\documentclass[letterpaper,11pt]{article}
\pdfoutput=1 

\usepackage{jheppub} 
                     
\usepackage{multirow}
\usepackage{verbatim}
\usepackage{pifont}

\usepackage{colortbl}
\usepackage{makecell}
\usepackage{hhline}

\newcommand{\be}{\begin{equation}}
\newcommand{\bea}{\begin{eqnarray}}
\newcommand{\beq}[1]{\begin{equation}\label{#1}}

\newcommand{\ee}{\end{equation}}
\newcommand{\eea}{\end{eqnarray}}
\newcommand{\eeq}{\end{equation}}
\newcommand{\lsim}{\!\mathrel{\hbox{\rlap{\lower.55ex \hbox{$\sim$}} \kern-.34em \raise.4ex \hbox{$<$}}}}
\newcommand{\gsim}{\!\mathrel{\hbox{\rlap{\lower.55ex \hbox{$\sim$}} \kern-.34em \raise.4ex \hbox{$>$}}}}

\newcommand{\vev}[1]{\langle #1 \rangle}
\newcommand{\abs}[1]{\left| #1 \right|}

\newcommand{\SU}{\mathrm{SU}}
\newcommand{\U}{\mathrm{U}}
\newcommand{\SO}{\mathrm{SO}}
\newcommand{\Sp}{\mathrm{Sp}}
\newcommand{\nlsm}{nl$\sigma$m}

\newcommand{\Sec}[1]{Sec.~\ref{#1}}
\newcommand{\Secs}[2]{Secs.~\ref{#1} and \ref{#2}}
\newcommand{\App}[1]{App.~\ref{#1}}
\newcommand{\Tab}[1]{Table~\ref{#1}}

\newcommand{\Fig}[1]{Fig.~\ref{#1}}

\newcommand{\Eq}[1]{Eq.~(\ref{#1})}
\newcommand{\Eqs}[2]{Eqs.~(\ref{#1}) and (\ref{#2})}
\newcommand{\Ref}[1]{Ref.~\cite{#1}}
\newcommand{\Refs}[1]{Refs.~\cite{#1}}

\begin{document} 

\begin{flushright}
MCTP-13-17 \\
MIT-CTP 4470
\end{flushright}
\vspace{0.2cm}

\title{Exotic Top Partners and Little Higgs}

\author[a,1]{John Kearney,\note{Corresponding author.}}
\author[a]{Aaron Pierce,}
\author[b]{and Jesse Thaler}
\affiliation[a]{Michigan Center for Theoretical Physics, Department of Physics, Ann Arbor, MI 48109, USA}
\affiliation[b]{Center for Theoretical Physics, Massachusetts Institute of Technology, Cambridge, MA 02139, USA}

\emailAdd{jkrny@umich.edu}
\emailAdd{atpierce@umich.edu}
\emailAdd{jthaler@mit.edu}

\abstract{
Little Higgs models often give rise to top partners beyond the minimal ones necessary for the cancellation of quadratic divergences.  We review how this occurs and discuss the phenomenology of these exotic states.  We emphasize the possible importance of new pseudo-Nambu-Goldstone bosons in top partner decays.  Indeed, cascade decays of exotic top partners may be the best way to discover these new bosons.  We illustrate these points with a new Little Higgs construction based on an $\SO(10)/\SO(5)^2$ coset structure, which fills a gap in the model building literature.  These observations motivate new search strategies for top partners at the LHC, including for final states with $b$-jets and a large multiplicity of electroweak bosons.}


\maketitle

\section{Introduction}

The discovery of the Higgs boson at the Large Hadron Collider (LHC) represents a turning point in our understanding of the electroweak theory.  Electroweak symmetry breaking is apparently accomplished via a scalar degree of freedom with properties not dissimilar from those of a Standard Model (SM) Higgs boson.  Despite this important advance, questions remain.  Foremost among these is whether the weak scale is natural and consequently whether there are additional structures beyond the SM that stabilize the Higgs boson mass against radiative corrections.

A scalar can be protected against large quadratically-divergent corrections to its (mass)$^2$ if it is composite \cite{Kaplan:1983fs,Kaplan:1983sm}.  In this case, radiative corrections are cut off at (or below) the compositeness scale $\Lambda \simeq 4 \pi f$, where $f$ is the relevant decay constant.  This approach was refined in the context of Little Higgs (LH) models \cite{ArkaniHamed:2002qx,ArkaniHamed:2002qy,Schmaltz:2005ky,Perelstein:2005ka}, which exhibit a parametric hierarchy between $f$ and the electroweak scale $v$.  These constructions mitigate precision electroweak tensions present in other strongly-coupled theories such as technicolor.  The separation between $v$ and $f$ is achieved in part by a collective breaking structure that allows the generation of a Higgs boson quartic coupling without the generation of a Higgs (mass)$^2$ parameter.  

Like any natural approach to electroweak symmetry breaking, LH models introduce partner particles for the top quark.  These top partners will be charged under $\SU(3)_C$, and so represent an appealing target for hadron collider searches.  There have been a number of phenomenological studies of fermionic top partners, with a primary focus on the minimal set needed for naturalness, the ``cancellons'' \cite{Perelstein:2003wd,Han:2003wu,Meade:2006dw} (see \Ref{DeSimone:2012fs,Buchkremer:2013bha} for recent reviews).  Cancellons are responsible for canceling quadratically-divergent contributions to the Higgs mass from SM top quark loops such that radiative corrections to the Higgs potential are at worst logarithmically divergent,
\begin{equation}
\label{eq:mhsqtuning}
\delta m_h^2 \simeq - \frac{3 y_t^2}{8 \pi^2} m_T^2 \log \left(\frac{\Lambda^2}{m_T^2}\right),
\end{equation}
where $m_T$ represents the cancellon mass and $y_t$ the top Yukawa coupling.  Minimizing tuning in the Higgs sector requires that the cancellons are as light as possible while still avoiding LHC constraints.  This suggests a mass scale for the top partners of $m_T \simeq f \simeq \mathcal{O}(\text{TeV})$, corresponding to a tuning of $\mathcal{O}(10-20\%)$ \cite{Berger:2012ec}.  Consequently, if naturalness is a reliable guide, top partners should soon be observed at the LHC.

In this paper, we study exotic top partners with novel decay patterns, motivated by the following logic:
\begin{itemize}
\item The Higgs boson emerges as a pseudo-Nambu-Goldstone boson (PNGB) from the coset space $G/H$.  It is possible, or perhaps even likely, that the top sector retains information about the underlying symmetry $G$ in spite of the fact that $G$ is ultimately broken by various gauge and Yukawa couplings.  In this case, top partners come in complete multiplets of $G$, although their masses are of course split by various spontaneous and explicit $G$-violating effects.  We will refer to these as ``long multiplets,'' which can contain exotic top partners in addition to the cancellons.
\item When exotic top partners appear in long multiplets, they may in fact be \emph{lighter} than the cancellons as a result of the collective breaking structure of LH theories  (see \Sec{subsec:exotictops}).  This simple observation has an important corollary: exotic top partners might be discovered \emph{prior} to the cancellon modes that actually regulate the Higgs potential.
\item Unlike minimal top partners---whose decays are likely dominated by the experimentally well-explored $T \rightarrow t h$, $T\rightarrow bW^+$, and $T\rightarrow t Z$ modes \cite{ATLAS:2012qe,Chatrchyan:2012vu,CMS:2012ab}---exotic top partners may well decay dominantly to other PNGBs besides the Higgs.   There are three kinds of PNGBs that are particularly well-motivated in the context of LH theories (see \Sec{subsec:exoticpngbs}):  ``quarticons,'' a second Higgs doublet, and extra ``uneaten'' goldstone bosons.  In particular, quarticons are those fields responsible for (collective) generation of the  Higgs boson quartic coupling, and their presence is required to maintain the hierarchy $v/f \ll 1$.  
\end{itemize}
Notably, direct production and observation of the additional PNGB scalars may be otherwise difficult.  Thus, it is possible that the best window into the structure of the LH theories may be via exotic decays of top partners into additional PNGBs.  These PNGBs will in turn dominantly decay to third-generation fermions and electroweak bosons to yield high-multiplicity final states at the LHC.

Some of the phenomenology we describe has previously appeared in the LH model building literature.  Indeed, specific examples are well-known to LH aficionados.  Here, we try to make the argument more general and extract a basic lesson: there is substantial motivation for top partner decays beyond the minimal ones.  Experimentally, this implies that searches for top partners should not be biased to exclusively look for $T \to t h, bW^+, t Z$ final states; decays like $T \to t hh$,  $T \to b W^+ Z$, and $T \to t b \bar{b}$ where $m_{b \bar{b}} \not= m_{h}$ are, among others, well-motivated possibilities in realistic LH constructions.  Furthermore, if a top partner is discovered, interpretation of its role with respect to naturalness must be made with care---it may or may not be the field responsible for stabilization of the weak scale.

The remainder of the paper organized as follows.  In \Sec{sec:structures}, we elaborate on the features of LH models outlined above, drawing on examples from the literature to illustrate the basic points.   We then summarize the main phenomenological consequences in \Sec{sec:exoticpheno}, discussing into which exotic scalars top partners might decay, and how the scalars themselves likely decay.  In \Sec{sec:so10modso5}, we present a new LH model based on an $\SO(10)/\SO(5)^2$ coset structure.  In addition to being a concrete illustration of the phenomenology of exotic top partners, the $\SO(10)/\SO(5)^2$ construction fills a ``missing box'' in the LH model building literature.  We present our conclusions in \Sec{sec:conclusions}.

\section{Lessons from Little Higgs Model Building}
\label{sec:structures}

In LH models, the Higgs boson emerges as a PNGB from the symmetry breaking pattern $G \rightarrow H$.  In this section, we first review the basic logic for why there must be a large global symmetry $G$ in LH theories, and argue why we expect top partners to come in complete $G$ multiplets (``long multiplets'').  Typically, if $G$ is larger than an $\SU(3)$, long multiplets will contain exotic top partners, which might actually be lighter than the cancellons.   We then discuss why LH theories likely contain additional PNGBs beyond a single Higgs doublet.  These ingredients---exotic top partners and extra PNGBs---will set the stage for the phenomenological discussion in \Sec{sec:exoticpheno}.

\subsection{Enlarged Global Symmetries and Long Multiplets}
\label{subsec:enlarged}

Why do we expect $G$ to be large?  In many cases, a large group $G$ is needed to generate a quartic coupling for the Higgs boson (see also \Sec{subsec:exoticpngbs}). Consider the case of the Simple Group LH \cite{Kaplan:2003uc}.  In the simplest (toy) model, an $\SU(3)$ gauge symmetry is broken down to the weak $\SU(2)_L$ by a pair of $\SU(3)$ triplet scalars, $\Phi_{1}$ and $\Phi_{2}$, with aligned vacuum expectation values (vevs).  The breaking pattern is $G/H = [\SU(3)/\SU(2)]^2$ to yield ten Goldstone bosons, and after five of them are eaten by the broken $\SU(3)$ gauge fields, the non-linear sigma model (\nlsm) fields are:
\begin{eqnarray}
\Phi_{1} = e^{i \Theta/f} 
\left(\begin{array}{c}
0 \\
0 \\
f \\
\end{array}
\right), \qquad
\Phi_{2} = e^{-i \Theta/f} 
\left(\begin{array}{c}
0 \\ 
0 \\
f
\end{array} 
\right),
\end{eqnarray}
with 
\begin{equation}
\Theta = \frac{1}{\sqrt{2}} \left( 
\begin{array}{ccc}
0 & 0 & \multirow{2}{*}{$h$} \\
0 & 0    \\
\multicolumn{2}{c}{h^{\dagger}} & 0 
\end{array}
\right) + \frac{\eta}{4} \left(\begin{array}{ccc} 1 & 0 & 0 \\ 0 & 1 & 0 \\ 0 & 0 & -2 \end{array}\right).
\end{equation}
Here, $h$ is the desired complex Higgs doublet, and $\eta$ is a real singlet.  Top partners can then be introduced as part of a complete $\SU(3)$ multiplet
\begin{equation}
\mathcal{Q} = \left( q, \chi_{u}\right)^{T},
\end{equation}
where $q$ is the SM quark doublet, $\chi_u$ is a cancellon field, and no exotic top partners are needed.  However, this toy model does not allow the generation of a Higgs quartic coupling without a large contribution to the Higgs (mass)$^{2}$.  The only non-trivial gauge-invariant that can be formed from the $\Phi_{i}$ fields is $\Phi_{1}^{\dagger} \Phi_2 = f^{2} + i f \eta - h^{\dagger} h$, and squaring this to yield a Higgs quartic coupling also introduces a problematic contribution to the Higgs (mass)$^{2}$.

To generate a quartic, the authors of \Ref{Kaplan:2003uc} therefore enlarge the symmetry to $G= \SU(4)^4$, broken down to $H=\SU(3)^4$ by \emph{four} scalar four-plets.   This approach works because it allows additional invariants for the quartic, some of which do not contribute to the Higgs boson mass.  A diagonal subgroup $\SU(4)_{V}$ is gauged, and is broken to the weak $\SU(2)_L$ because of misalignment of the four-plet vevs.  The consequence of having an underlying $\SU(4)$ structure is that the fundamental building block of the top sector is a four-plet
\begin{equation}
\label{eq:fourplet}
\mathcal{Q} = \left( q, \chi_{u1}, \chi_{u2}\right)^{T}.
\end{equation}
As we will discuss more in \Sec{subsec:exotictops}, $\chi_{u1}$ is a cancellon field but $\chi_{u2}$ is an exotic top partner whose presence is only necessitated by the enlarged group $G$.  More generally, we would expect that some top sector fields transform as long multiplets under whatever extended global symmetry was used to achieve a Higgs quartic coupling.

There are LH scenarios where, instead of enlarging $G$ from $\SU(3)$ to $\SU(4)$ to generate a Higgs quartic coupling, one enlarges $\SU(3)$ to $\SU(3)^n$.  This is the approach taken in the Minimal Moose LH models \cite{ArkaniHamed:2002qx,Gregoire:2002ra}.  Since $\SU(3)^n$ is a product group, the top partners need only transform under one of the $\SU(3)$'s and no exotics are necessary.    That said, minimality is not a principle of nature, and moose models with more general $G^n$ symmetries are certainly plausible.  Such models would contain exotics if the top partners came in long multiplets.

Another class of LH model featuring large $G$ is based on the ``Littlest Higgs'' structure.  In this case, $G$ must be large enough to contain two subgroups that act non-linearly on the Higgs in order to achieve collective symmetry breaking.  For instance, in the Littlest Higgs \cite{ArkaniHamed:2002qy}, two $\SU(3)$ symmetries protect the Higgs, requiring $G$ be (at least) $\SU(5)$---we will elaborate on the required size of $G$ in models of this type in \Sec{sec:so10modso5}.  Imposing that fermions come in long multiplets (e.g.~complete multiplets of $\SU(5)$) leads to the presence of additional non-cancellon top partners.
In the case of \Refs{Katz:2003sn,Thaler:2005en}, which implement the ``Littlest Higgs'' construction of \Ref{ArkaniHamed:2002qy} with long multiplets, these exotic top partners are the $p$-fields.
As we will discuss in more detail below, if these additional fields do not participate in soft $G$-breaking then they can in fact be lighter than the cancellons.  

It should now be clear that the $G$ of LH constructions can be large (larger than $\SU(3)$), and how in principle $G$ might be reflected in the fermion structure of the theory.  However, this by itself does not tell us whether fermions are indeed required to come in complete $G$ multiplets.
In the case of the Simple Group models, complete $G$ multiplets are required by gauge invariance.  More generally, though, most LH models take the form of $G/H$ with only a subgroup $F \subset G$ gauged.  An example is the Littlest Higgs \cite{ArkaniHamed:2002qy} where $G/H = \SU(5)/\SO(5)$ and $F = [\SU(2) \times \U(1)]^2$.   In this case, theoretical consistency of the gauge symmetry only requires top partners to come in complete $F$ (not $G$) multiplets.  Alternatively, taking a low-energy perspective, top partners need only transform under the unbroken symmetry $H$ since one can always use the CCWZ formalism \cite{Callan:1969sn,Coleman:1969sm} to lift an $H$ multiplet to a $G$ multiplet.  So while we have good reasons to expect top partners to come in $F$ multiplets or $H$ multiplets, it is not clear why one should expect them to come in $G$ multiplets. 

Our logic for complete $G$ multiplets is as follows.  If the Higgs is a PNGB arising from the breaking of $G \rightarrow H$ by strong dynamics, then $G$ should be a good symmetry of the strong dynamics at high energies, only broken by (small) couplings between composite and elementary operators.  It is therefore a distinct possibility (though not a requirement) that the strong dynamics generates only $G$-invariant couplings of fermions to the Higgs field.  In that case, some fermions (corresponding to the ones arising from the strong dynamics) must have quantum numbers corresponding to complete multiplets of $G$ in order to produce those invariant couplings.

We expect these complete $G$ multiplets to be subsequently split into smaller multiplets as a result of both spontaneous and explicit $G$-breaking effects.  At minimum, the spontaneous breaking $G \rightarrow H$ will result in complete $G$ multiplets factorizing into smaller $H$ multiplets with $\mathcal{O}(1)$ mass splittings.  In realistic models, $G$ must also be explicitly broken in the gauge and fermion sectors such that the fields parameterizing the coset space $G/H$ are pseudo (rather than exact) NGBs.  In the gauge sector, gauging $F$ further (explicitly) breaks the degeneracy of the $G$ multiplets, such that only multiplets of $F \cap H$ will be exactly degenerate.  In the fermion sector, explicit breaking can always be made ``soft,'' accomplished via mass mixing of composite fermions with fundamental fields unrelated to the strong dynamics, leading to further mass splittings within the $G$ multiplets without generating quadratic divergences in the Higgs potential.

\begin{figure}
\centering
\includegraphics[width=0.35\textwidth]{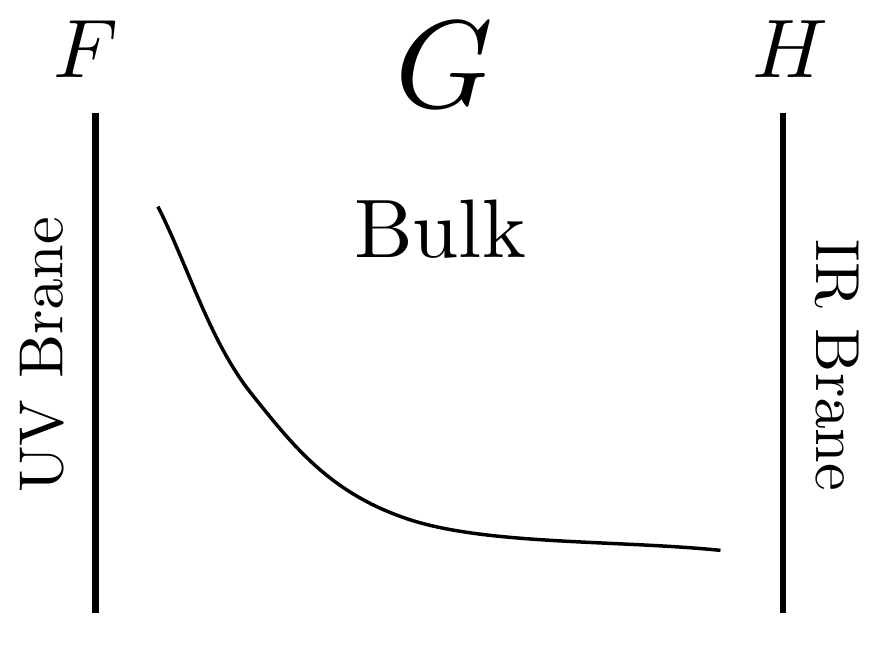}
\caption{A Little Higgs model realized in a slice of AdS$_5$, adapted from \Ref{Thaler:2005en}.  If top partners are in the bulk, then they must come in complete $G$ multiplets.  Only if the top partners are entirely localized on the UV (IR) branes would one expect them to only be in $F$ ($H$) multiplets.  UV and IR boundary conditions for bulk top partners will result in complete $G$ multiplets being factorized into $F \cap H$ multiplets, leading to multiplets with $\mathcal{O}(1)$ mass splittings.}
\label{fig:AdSBulkPicture}
\end{figure}

The AdS dual of this strongly-coupled logic was realized in \Ref{Thaler:2005en} (see also \Refs{Contino:2003ve,Agashe:2004rs,Thaler:2005kr}).  There, a LH model was constructed on a slice of AdS$_5$ bounded by a UV brane and an IR brane, shown in \Fig{fig:AdSBulkPicture}.  The requirement that $G$ be a good symmetry of the strong dynamics implies that $G$ is a gauge symmetry in the bulk.  In particular, bulk fermions (including the top partners) are introduced in complete $G$ multiplets.  The breaking $G \rightarrow H$ occurs on the IR brane, making it intuitively obvious that the PNGBs correspond to composite states of the strong dynamics.  All explicit $G$-breaking effects (such as the gauging of the $F$ subgroup) are localized on the UV brane, ensuring that the PNGB fields are effectively ``shielded'' from $G$-breaking.  These boundary effects result in $\mathcal{O}(1)$ mass splittings within fermion $G$ multiplets even though, from the point-of-view of the bulk, $G$ is still the ``correct'' underlying symmetry.\footnote{In the context of vector mesons in QCD, the $\rho$ and $a$ states can be considered to fill out a complete multiplet of $\SU(2)_L \times \SU(2)_R$, though of course they are not degenerate after chiral symmetry breaking.}  This AdS construction is an explicit example that avoids the ``hidden fine-tuning'' warned about in \Ref{Grinstein:2011qm}, since the bulk $G$ gauge symmetry ensures that $G$ is indeed a ``good enough'' symmetry, despite the explicit $G$-breaking boundary conditions.\footnote{An analogous feature is familiar in the SM, where the $\pi^\pm/\pi^0$ mass splitting from photon loops is not logarithmically sensitive to high scale physics because QCD dynamics preserves isospin.}

\subsection{Exotic Top Partners}
\label{subsec:exotictops}

Having established the likelihood of both enlarged global symmetries $G$ and complete $G$ top multiplets, we now turn to the consequences of having long  multiplets.  A key question is how the exotic top partners obtain mass.  One possibility is that the exotic partners experience large explicit $G$-breaking and are simply lifted out of the low energy spectrum.  More interesting for our purposes is if the exotic top partners get mass from the spontaneous $G \to H$ breaking dynamics, in which case exotic top partners should be near the weak scale.  We will look at two scenarios: one where the exotic top partner mass is unrelated to the generation of the SM top Yukawa, and one where it is related.

As an example of the first case, we return to the Simple Group LH.  In \Ref{Kaplan:2003uc}, the quark four-plet in \Eq{eq:fourplet} marries up with quark singlets as
\begin{equation}
\label{eq:SGyukawa}
\mathcal{L} \supset y_{1} \Phi_{1}^{\dagger}  \mathcal{Q}   \chi_1^{c}  + y_{2} \Phi_{2}^{\dagger}  \mathcal{Q}   \chi_2^{c} + y_{3} \Psi_{1}^{\dagger} \mathcal{Q} \chi_3^c ,
\end{equation}
where the four scalar four-plets get vevs $\langle \Phi_{1} \rangle = ( 0, 0,  f_{1}, 0 )^{T}$,   $\langle \Phi_{2} \rangle = ( 0, 0,  f_{2}, 0 )^{T}$, $\langle \Psi_{1} \rangle = ( 0, 0,  0,  f_{3} )^{T}$, and $\langle \Psi_{2} \rangle = ( 0, 0,  0, f_{4} )^{T}$.
Consequently, prior to electroweak symmetry breaking, the top Yukawa is
\begin{equation}
y_{t} = \frac{y_1 y_2}{\sqrt{2}} \sqrt{\frac{f_{1}^2 + f_{2}^2}{y_{1}^2 f_{1}^2 + y_{2}^2 f_{2}^2}},
\end{equation}
and the spectrum contains two vector-like electroweak singlets with masses
\begin{align}
m_{T} & = \sqrt{y_{1}^{2} f_{1}^2 + y_{2}^2 f_2^2} & & \rm{(cancellon),}\\
m_{X} & = y_{3} f_{3} & & \rm{(exotic).}
\end{align}
Electroweak symmetry breaking will induce mixing between the charge-$2/3$ quarks, producing corrections to these expressions that are suppressed by powers of $v/f$.  As anticipated, the $y_{3}$ term responsible for giving mass to the exotic does not participate in the generation of the top Yukawa.  Correspondingly, the existence of this exotic $X$ is not necessary to cancel quadratic divergences in this model and only appears because $\mathcal{Q}$ is a complete multiplet of $\SU(4)$, which is split by the spontaneous $G$-breaking into $\SU(2)_L$ multiplets.\footnote{Similarly, the field  $\chi_3^c$ is only needed to marry off $\chi_{u2}$.} Precisely because it plays no role in the the Yukawa coupling generation, $y_{3}$ is a completely free parameter untied to naturalness, so the exotic top partner can be either lighter or heavier than the cancellon field.  

In alternative top sectors, though, the exotics can indeed get masses from the same mechanism that generates the top Yukawa coupling.  Take the  ``Littlest Higgs'' type structure \cite{ArkaniHamed:2002qy} where the breaking $G/H$ is achieved via a $\Sigma$ field that transforms schematically as\footnote{Here, tilde represents the operation of sending all broken generators $X \to -X$.  If this operation is possible, then the resulting $G/H$ is called a symmetric space.}
\be
\Sigma \to G \Sigma \widetilde{G}^{-1}
\ee
and gets a vev $\vev{\Sigma} = \mathbf{1}$ in some basis.  We consider Yukawa couplings of the form first introduced in \Ref{Katz:2003sn} and subsequently studied in \Refs{Thaler:2005en,Vecchi:2013bja}.  The $G$-invariant Yukawa couplings generated by the strong dynamics are
\begin{equation}
\label{eq:GinvYukawa}
\mathcal{L}^{G-\text{inv}}_{\text{Yuk}} = - y_1 f \mathcal{Q} \Sigma \mathcal{Q}^c + \text{h.c.}
\end{equation}
where $\mathcal{Q}, \mathcal{Q}^c$ are complete $G$ multiplets.  This coupling exhibits an expanded symmetry $G_L \times G_R$ under which, schematically,
\be
\Sigma \rightarrow L \Sigma R^{-1}, \qquad \mathcal{Q} \rightarrow \mathcal{Q} L^{-1}, \qquad \mathcal{Q}^c \rightarrow R \mathcal{Q}^c.
\ee
The vev $\vev{\Sigma}$ then breaks $G_L \times G_R \rightarrow H_V$, the vector combination of the subgroups $H_{L, R} \subset G_{L, R}$.  To recover the SM as a low-energy theory, an additional $\SU(2)_L$ doublet $q$ and an $\SU(2)_L$ singlet $u^c$ are introduced with mass terms
\begin{equation}
\label{eq:softGbreaking}
\mathcal{L}^{\text{soft}}_{\text{mass}} = - y_2 f U u^c - y_3 f q Q^c + \text{h.c.}
\end{equation}
where $U$ ($Q^c$) is the component of $\mathcal{Q}$ ($\mathcal{Q}^c$) with the appropriate electroweak quantum numbers to form the above mass terms.

Mixing between the ``composite'' fermions in $\mathcal{Q}, \mathcal{Q}^c$ and the ``fundamental'' fields $q, u^c$ gives rise to both SM top fields and cancellons responsible for regulating the Higgs potential.  In fact, because there are effectively two cancellon fields ($T$ and $Q_3$ defined below), the top sector contribution to the radiatively-induced Higgs potential is finite and calculable at one-loop.\footnote{\label{footnote:modtop}A variation on this top sector was realized in \Ref{Schmaltz:2010ac}, in which the \nlsm\ field $\Sigma$ appears in all three $y_i$ terms.  This allows the top Yukawa to be generated with lighter cancellon masses.}  Said another way, this fermion structure exhibits ``triple protection'', such that $y_1$, $y_2$ and $y_3$ must all be non-zero for a top Yukawa coupling to be generated.\footnote{For example, the $y_2$ term explicitly breaks $G_L$ while preserving $G_R$, so for $y_1, y_2 \neq 0$ but $y_3 = 0$, the breaking of the remaining exact global symmetry $G_R \rightarrow H_V$ by $\vev{\Sigma}$ ensures that the Higgs is an exact NGB.  An analogous story holds for $y_2 = 0$.  For $y_1 = 0$, $\Sigma$ decouples from the top sector entirely.}  But beyond the cancellons, this top sector requires exotic top partners because long top multiplets are needed to produce the $G$-invariant coupling in \Eq{eq:GinvYukawa}.

Explicitly, let $Q^{(c)}, U^{(c)}, X^{(c)}$ denote the components of $\mathcal{Q}^{(c)}$, where $X^{(c)}$ are the set of all exotic fields.  Expanding \Eqs{eq:GinvYukawa}{eq:softGbreaking}, one finds
\begin{align}
\mathcal{L}^{G-\text{inv}}_{\text{Yuk}} + \mathcal{L}^{\text{soft}}_{\text{mass}} &= - y_1 f \left(Q Q^c + U U^c + X X^c - Q \frac{h}{f} U^c + \ldots\right) \\
&\qquad ~ - y_2 f U u^c - y_3 f q Q^c + \text{h.c.}  + \ldots \nonumber,
\end{align}
where the dots indicate higher order terms and terms involving PNGBs besides the Higgs doublet $h$.   Prior to electroweak symmetry breaking, the mass eigenstates are
\begin{align}
\label{eq:topfieldredef}
Q_3 & \equiv \frac{y_1 Q + y_3 q}{\sqrt{y_1^2 + y_3^2}}, & 
q_3 & \equiv \frac{y_3 Q - y_1 q}{\sqrt{y_1^2 + y_3^2}}, &
T^c & \equiv \frac{y_1 U^c + y_2 u^c}{\sqrt{y_1^2 + y_2^2}}, & 
t^c & \equiv \frac{y_2 U^c - y_1 u^c}{\sqrt{y_1^2 + y_2^2}},
\end{align}
such that
\begin{align}
\label{eq:GinvWithX}
\mathcal{L}^{G-\text{inv}}_{\text{Yuk}} + \mathcal{L}^{\text{soft}}_{\text{mass}} &= - \sqrt{y_1^2 + y_2^2} f U T^c - \sqrt{y_1^2 + y_3^2} f Q_3 Q^c - y_1 f X X^c \\
&\qquad ~+ \frac{y_1 y_2 y_3}{\sqrt{y_1^2 + y_2^2} \sqrt{y_1^2 + y_3^2}} q_3 h t^c + \ldots \nonumber,
\end{align}
To leading order, we can identify $q_3$ and $t^c$ as the SM top doublet and singlet, respectively, with a top Yukawa given by
\begin{equation}
y_t = \frac{y_1 y_2 y_3}{\sqrt{y_1^2 + y_2^2} \sqrt{y_1^2 + y_3^2}}.
\end{equation}
As expected from the symmetry argument given above, this vanishes unless all $y_i \neq 0$.  There are also a number of vector-like fermions:
\begin{align}
m_{T} & = \sqrt{y_1^2 + y_2^2} f && \text{(singlet cancellon)}, \label{eq:Ucancellon}\\
m_{Q_3} & = \sqrt{y_1^2 + y_3^2} f &&  \text{(doublet cancellon)},\\
m_{X} &= y_1 f && \rm{(exotics)}.
\end{align}
Because of the expanded global symmetry in the top sector, spontaneous $G$-breaking leaves this long multiplet degenerate, with mass splittings arising only from the explicit breaking by $y_2$ and $y_3$.  The cancellons are responsible for the collective breaking of the Higgs shift symmetries.  However, $G$-invariance also gives rise to vector-like non-cancellon fields $X, X^c$ which (in this construction) are in fact \emph{lighter} than the cancellons and thus more readily accessible at colliders.

An interesting corollary is that, if we do not observe exotic top partners at a particular mass, then the cancellons may be even heavier, thereby increasing naturalness tensions in LH theories.\footnote{One could introduce additional (spontaneous or explicit) $G$-breaking effects to lift the mass of the exotics, though this is at odds with our philosophy of trying to maintain $G$ as a good global symmetry of the strong dynamics to the extent possible.}  That said, the ``triple protection'' described above modifies \Eq{eq:mhsqtuning} such that the logarithmic divergence is removed \cite{Katz:2003sn},
\begin{equation}
\delta m_h^2 \simeq -\frac{3 y_t^2}{8 \pi^2} \frac{m_T^2 m_{Q_3}^2}{m_T^2 - m_{Q_3}^2} \log\left(\frac{m_T^2}{m_{Q_3}^2}\right).
\end{equation}
As a result, even for exotic top partners with $m_X \gsim 600 \text{ GeV}$ (in excess of current LHC limits \cite{ATLAS:2012qe, Chatrchyan:2012vu,CMS:2012ab,Rao:2012gf}), tuning is only $\mathcal{O}(50\%)$ provided the cancellons are not substantially heavier.  Bounds on $m_X$ can increase dramatically before these models start to exhibit significant fine-tuning; for $y_1 = y_2 = y_3$, tuning is $\mathcal{O}(10\%)$ for $f \gsim 700 \text{ GeV}$, corresponding to $m_X \gsim 1.4 \text{ TeV}$ and $m_T = m_{Q_3} \gsim 2 \text{ TeV}$.

\subsection{Additional PNGBs}
\label{subsec:exoticpngbs}

Just as enlarged $G$ symmetries can necessitate exotic top partners in long multiplets, large $G/H$ coset spaces can give rise to additional PNGBs beyond just a single Higgs doublet.  There are three particularly well-motivated additions: quarticons, a second Higgs doublet, and extra uneaten goldstone bosons.

At minimum, a successful LH theory must include quarticons to achieve a collective Higgs quartic coupling \cite{Schmaltz:2008vd}.  In order to protect the Higgs doublet(s) from radiative (mass)$^2$ corrections yet still generate a quartic interaction, the potential must contain operators (included by hand or induced radiatively) of the schematic form
\begin{equation}
\label{eq:quarticonL}
V \supset \lambda_+ f^2 \abs{\varphi + \frac{h^2}{f}}^2 + \lambda_- f^2 \abs{\varphi - \frac{h^2}{f}}^2,
\end{equation}
where $\varphi$ are the quarticons, and $h$ is a stand-in for one or more Higgs doublets.  Each term in \Eq{eq:quarticonL} preserves one of two shift symmetries acting on the Higgs:
\begin{align}
\label{eq:quarticonshifts1}
\delta_\epsilon h = \epsilon, & \qquad \delta_\epsilon \varphi = - \frac{\epsilon h + h \epsilon}{f}, \\
\label{eq:quarticonshifts2}
\delta_\eta h = \eta, & \qquad  \delta_\eta \varphi = \frac{\eta h + h \eta}{f}.
\end{align}
If only $\lambda_+$ or $\lambda_-$ is present, the potential respects one of the shift symmetries and the Higgs is an exact NGB.\footnote{The apparent Higgs interactions can be  removed by a field redefinition $\varphi \rightarrow \varphi \pm \frac{h^2}{f}$.}  When both operators are present, the quarticon gets a mass\footnote{There would be a factor of two if the quarticon were real instead of complex.}
\be
\label{eq:quarticonmass}
m_\varphi = f \sqrt{\lambda_+ + \lambda_-},
\ee
and integrating out $\varphi$ generates a quartic coupling $\lambda h^4$ with
\begin{equation}
\label{eq:quartic}
\lambda = \frac{4 \lambda_+ \lambda_-}{\lambda_+ + \lambda_-}.
\end{equation}
Because the quarticons transform under the same shift symmetries that protect the Higgs, they must be PNGBs from the coset space $G/H$.

Following \Ref{Schmaltz:2008vd}, one can determine the electroweak quantum numbers of the quarticons from the $\SU(2)_L \times \U(1)_Y$ invariance of \Eq{eq:quarticonL}.  In the case of a single Higgs doublet, $\varphi$ must be a (complex or real) electroweak triplet.\footnote{\label{footnote:whyDangerous}Though it would have the right $\SU(2)_L \times \U(1)_Y$ quantum numbers, a real singlet $\varphi$ is ``dangerous'' \cite{Schmaltz:2008vd}.  Since $\varphi$ has no non-trivial quantum numbers, a quadratically-divergent $\varphi$ tadpole---which by the shift symmetries in \Eqs{eq:quarticonshifts1}{eq:quarticonshifts2} is necessarily accompanied by an undesirable quadratically-divergent Higgs mass---will arise at one-loop.}  Without a symmetry such as $T$-parity \cite{Cheng:2003ju,Cheng:2004yc,Low:2004xc}, though, a triplet quarticon will generically get a vev, in tension with precision constraints on the electroweak $\rho$ parameter.  One way to avoid electroweak precision constraints is to extend $G$ such that $H$ includes a custodial symmetry as in \Refs{Chang:2003un,Chang:2003zn}.  If one accepts the philosophy of \Sec{subsec:exotictops}, this would require top partners to come in complete multiplets of $\SU(2)_L \times \SU(2)_R \cong \SO(4)$---in addition to the proliferation of non-cancellon top partners, this implies the presence of an exotic charge-$5/3$ quark \cite{Agashe:2006at}.

Alternatively (or additionally, as in the $\SO(10)/\SO(5)^2$ model in \Sec{sec:so10modso5}), $G/H$ could be expanding to include a second Higgs doublet.  This permits (real and complex) singlet quarticons,\footnote{\label{footnote:whytwoHiggs}The reason these singlets do not suffer from the dangerous singlet pathology is that $H_u H_d$ is not required to be a singlet under all symmetries.  Thus, singlet tadpoles and the associated Higgs (mass)$^2$ can be prohibited by a parity in the case of a real singlet \cite{Schmaltz:2010ac} or a $\U(1)_{\text{PQ}}$ symmetry \cite{Peccei:1977hh,Peccei:1977ur} in the case of a complex singlet \cite{Low:2002ws}.} alleviating the issue of triplet vevs without resorting to $T$-parity.  If $\tan \beta \equiv v_u/v_d$ is close to one, then LH models with two Higgs doublets are in good agreement with electroweak precision measurements \cite{Gregoire:2003kr}.  Alternately, one can introduce a custodial symmetry on top of the two Higgs doublet structure to have more general values of $\tan \beta$ \cite{Schmaltz:2010ac}.  In either case, precision measurements give strong motivation to consider LH models with a second Higgs doublet.  A second Higgs doublet may also be present simply due to an enlarged global symmetry.  For instance, in the Simple Group LH \cite{Kaplan:2003uc}, enlarging $G$ from $\SU(3)^2$ to $\SU(4)^4$ in order to generate the quartic introduces both a second Higgs doublet and (complex) singlet quarticons.

Finally, ``uneaten'' PNGBs may be present due to ``modular breaking'' of the gauge groups as in \Ref{Schmaltz:2010ac}.  In LH models, the breaking of the global symmetry $G \rightarrow H$ is generally accompanied by breaking of the gauged subgroup $F \rightarrow \SU(2)_L \times \U(1)_Y$.  Some of the NGBs from $G/H$ are eaten to produce massive $W^\prime$ gauge bosons with schematic mass
\begin{equation}
m_{W^\prime} \simeq g f
\end{equation}
where $g$ is a gauge coupling.  However, experimental constraints---particularly from  four-fermion operators generated by integrating out the $W^\prime$ bosons---generally require these bosons to be quite heavy, \cite{Hewett:2002px,Csaki:2002qg,Csaki:2003si,Gregoire:2003kr,Perelstein:2003wd,Kilian:2003xt}.  Heavier $W'$ bosons require larger $f$, which in turn requires heavier top partners, at odds with electroweak naturalness.

The setup of modular breaking involves a second \nlsm\ field $\Delta$ that breaks $F \rightarrow \SU(2)_L \times \U(1)_Y$ at a scale $V > f$.  As a result, the heavy gauge boson get mass $m_{W^\prime} \simeq g V$, but the scale $f$ can remain somewhat lower.   In this case, a set of PNGBs remains uneaten (corresponding to a linear combination of the would-be eaten fields in $\Sigma$ and those in $\Delta$).  This is analogous to the situation in the SM where $\SU(2)_L \times \U(1)_Y$ is broken both by a quark condensate and, at a significantly higher scale, by the Higgs field, yielding three light uneaten pions at energies below $\Lambda_{\text{QCD}}$.  In the LH case, the uneaten PNGBs acquire mass of order (at least) $g_{\text{EW}} V/(4 \pi)$ through loops of gauge bosons, so they may be quite light and therefore produced in top partner decays.\footnote{While uneaten PNGBs appeared in \Ref{Schmaltz:2010ac}, they largely decoupled from the phenomenology due to the unique form of the top sector alluded to in footnote~\ref{footnote:modtop}.  With the standard top Yukawa structure in \Eq{eq:GinvYukawa}, there are indeed couplings between the top partners, SM tops, and the uneaten PNGBs.}

\section{Phenomenological Consequences}
\label{sec:exoticpheno}

As argued above, there are compelling theoretical arguments to expect LH models to contain (1) top partners beyond the cancellons responsible for regulating the Higgs potential and (2) additional PNGBs beyond a single Higgs doublet, including quarticons, a second Higgs doublet, and uneaten PNGBs.  In this section, we show how these ingredients give rise to interesting phenomenology beyond the vanilla top partner decays $T \rightarrow bW^+, tZ, th$ usually considered.  Especially for models with top sectors of the form in \Eqs{eq:GinvYukawa}{eq:softGbreaking}, there can be a host of top partners (both cancellons and lighter non-cancellons) within reach of the LHC, motivating new search strategies.  We expect that top partners will be dominantly pair produced via QCD processes, though single production may also be important \cite{Han:2003wu,Buchkremer:2013bha}.  

One well-appreciated possibility is that complete $G$ multiplets may include top partners with exotic charges.  As mentioned above, if the coset space $G/H$ exhibits a custodial symmetry, then the SM quark doublet may be part of a custodial doublet as well, implying the existence of an exotic charge-5/3 quark \cite{Agashe:2006at}.  This quark can decay as $X^{5/3} \rightarrow t W^+ \rightarrow b W^+ W^+$, yielding exotic same-sign dilepton signatures.  Such signals have been searched for at the LHC and limits of $m_{X^{5/3}} \gsim 650-700 \text{ GeV}$ have been placed \cite{ATLAS:2012hpa,CMS:2012cya}.

More exotic phenomenology can arise from interactions between exotic top partners and additional PNGBs.   With top partners, quarticons, Higgs doublets, and uneaten PNGBs all transforming under $G$, $G$-invariance will generate renormalizable couplings between exotic top partners, additional PNGBs, and SM fermions.  Such couplings are present, for example, in the models of \Refs{Katz:2003sn,Kaplan:2003uc,Thaler:2005en,Vecchi:2013bja,Schmaltz:2010ac} (all of which, at least approximately, follow the $G$-invariant philosophy).  These couplings can permit novel top partner decays of the schematic form (with $Q$ denoting a top partner and $q$ a third generation SM quark):
\begin{itemize} 
\item{$Q \rightarrow \sigma q$, with $\sigma$ an ``exotic'' (non-Higgs) PNGB};
\item{$Q \rightarrow H_{d} q$, which would give rise to final states involving the components of the second Higgs doublet, $H^0$, $A^0$, and $H^\pm$ (although mixing between $H_{u}$ and $H_{d}$ would lead to the usual $Q \rightarrow b W^+, tZ, th$ decays as well).}
\end{itemize}
As direct electroweak production of PNGB states in hadron colliders would be limited, exotic top partner decays may provide the best avenue for discovering these additional bosons at the LHC.

The first important question is whether these decays are kinematically allowed.  It seems likely that the answer is yes.  Because the top Yukawa $y_t \approx 1$, this suggests that couplings in the Yukawa sector are $y_i \simeq \mathcal{O}(1)$, implying top partners with masses $m \gsim f$.  Meanwhile, PNGB masses tend to be related to electroweak gauge or quartic couplings, and are therefore lighter.  For quarticons in particular, recent measurements of $m_h \approx 125 \text{ GeV}$ suggest a quartic $\lambda \approx \frac{1}{4}$.  \Eqs{eq:quarticonmass}{eq:quartic} suggest $\lambda_\pm \simeq \lambda \Rightarrow m_{\varphi} \simeq \sqrt{\lambda_\pm} f \lsim f$, making quarticons light enough to be produced in top partner decays.  

One might then ask whether interactions involving top partners and additional PNGBs are suppressed by $v/f$ relative to expected decays to $th$, $tZ$, and $bW^+$.  While this can be the case, it is generally not for all top partners.  For instance, in the Simple Group model \cite{Kaplan:2003uc} (see \Eq{eq:SGyukawa}), the exotic $X \rightarrow q H_d$ decay can arise directly from the $\chi_3 H_d^\dagger q$ Yukawa coupling, such that which $X$ decay modes dominate will depend on the interplay between $\tan \beta$, $v/f$ corrections, and phase space suppression.  Recently, \Ref{Godfrey:2012tf} analyzed top partner decays in the Bestest LH \cite{Schmaltz:2010ac}, finding that decays to the second Higgs doublet can be substantial for certain parameter choices.  In \Sec{subsec:so10topsector}, we will present further examples of top partners with leading-order decays to SM top fields and additional exotic PNGBs in the context of an $\SO(10)/\SO(5)^2$ LH model.  So, while exotic decay branching ratios will be model-dependent, such decays are well-motivated and significant in sizable regions of parameter space.

The appropriate search strategy for uncovering these exotic top partner decays depends on how the PNGBs themselves decay (see \Tab{tab:decaytopologies}).  Second Higgs doublet states ($H^0, A^0$, and $H^\pm$) likely decay predominantly to third-generation quarks (assuming we are in a quasi-decoupling regime).
The situation is somewhat more complicated for quarticons.  As they have the correct electroweak quantum numbers to couple to pairs of Higgs doublets, quarticons cannot form $\SU(2)_L$-invariant Yukawa couplings with the SM top fields $q_3$ and $t^c$.  Thus, quarticon decays to third-generation quarks, such as $\varphi^+ \rightarrow t \bar{b}$ or $\varphi^0 \rightarrow t \bar{t}$, must be suppressed by powers of $v/f$.  By contrast, couplings of the form 
\begin{equation}
(\lambda_+ - \lambda_-) f \varphi h_1 h_2
\end{equation}
necessarily arise from the collective quartic structure of \Eq{eq:quarticonL} with two Higgs doublets, and these permit quarticons to decay to longitudinal gauge bosons and Higgs bosons at leading order (assuming no symmetry forcing $\lambda_+ = \lambda_-$, such as $T$-parity).  In addition, decays such as $\varphi^0 \rightarrow t \bar{t}$ or $\varphi^0 \rightarrow H^+ H^-$ suffer from phase space suppression relative to $\varphi^0 \rightarrow W^+ W^-, Z Z,$ and $h h$.\footnote{While we have written the decays as for a neutral quarticon $\varphi^0$ in order to be explicit, this should also be true for charged quarticons, which can decay via, e.g., $\varphi^+ \rightarrow W^+ Z$ or $\varphi^{++} \rightarrow W^+ W^+$.}  Thus, it seems likely that quarticons chiefly decay into pairs of light Higgs and electroweak gauge bosons, though the exact branching ratios vary with masses and couplings, and one channel does not clearly dominate over the entire parameter space.  Uneaten PNGBs have the same quantum numbers as gauge bosons, so also exhibit $v/f$ suppression in their couplings to SM quark pairs.  However, since they could have been eaten by the $W^\prime$ gauge bosons were it not for the modular breaking structure, their couplings to electroweak bosons tend to be further suppressed.  Consequently, like second Higgs doublet states, uneaten PNGBs are expected to decay to third-generation fermions and do not require alternative search strategies beyond those developed for the two Higgs doublet model (2HDM).

\begin{table}
\centering
\renewcommand{\arraystretch}{1.2}
\begin{tabular}{|| c c l l l || c | c c l l || }
\hline
\multicolumn{5}{|| c||}{\textbf{Exotic Top Partner Decays}} & \multicolumn{5}{c ||}{\textbf{Expected PNGB Decays}} \\
\hline
$X^{5/3}$ & $\rightarrow$ & $t H^+,$ & $t \varphi^+$ & & \multirow{3}{*}{Quarticons} & $\varphi^{++}$ & $\rightarrow$ & $V^+ V^+$ & \\
$X^{5/3}$ & $\rightarrow$ & $b \varphi^{++}$ & & & & $\varphi^{+}$ & $\rightarrow$ & $V^+ V^0$ & \\
$T^{2/3}$ & $\rightarrow$ & $t H^0,$ & $t A^0,$ & $t \varphi^0$ & & $\varphi^{0}$ & $\rightarrow$ & $V^+ V^-,$ & $V^0 V^0$ \\ \cline{6-10}
$T^{2/3}$ & $\rightarrow$ & $b H^+,$ & $b \varphi^+$ & & \multirow{3}{*}{2\textsuperscript{nd} Higgs Doublet} & $H^+$ & $\rightarrow$ & $t \bar{b}$ & \\
$B^{-1/3}$ & $\rightarrow$ & $b H^0,$ & $b A^0, $ & $b \varphi^0$ & & $H^0$ & $\rightarrow$ & $t \bar{t}$, & $b \bar{b}$ \\
$B^{-1/3}$ & $\rightarrow$ & $t H^-, $ & $t \varphi^-$ & &  {\footnotesize (also uneaten PNGBs)} & $A^0$ & $\rightarrow$ & $t \bar{t}$, & $b \bar{b}$ \\
\hline
\end{tabular}
\caption{\label{tab:decaytopologies}Exotic top partner decays, where $Q^q$ and $\varphi^q$ denote top partners and quarticons of charge $q$, respectively.  $V$ represents either an electroweak gauge boson ($W^\pm$ or $Z$) or a scalar Higgs boson ($h$, $H^0$, $A^0$, or $H^\pm$).  Due to phase space suppression, quarticon decays to electroweak or light Higgs boson pairs are expected to dominate (i.e.~modes with $V = h, W^\pm, Z$).  Uneaten PNGBs are expected to exhibit similar decays to second Higgs doublet states.  In addition to these decay modes, there are potentially additional modes with decays widths suppressed by $v^2/f^2$.}
\end{table}

Based on these observations, searches for exotic  top partners decaying to third-generation- and electroweak-boson-rich final states are particularly well-motivated, and should be pursued at the LHC.  \Tab{tab:decaytopologies} lists possible exotic top partner decays to  third-generation quarks plus additional PNGBs.  Also shown are the likely dominant decay modes of the PNGBs.  Note that there is a wide variety of possibilities, and that a number of the decay chains give rise to different kinematics or even radically different final states to those usually assumed for top partner decays.  

\begin{figure}
\centering
\includegraphics[width=0.8\textwidth]{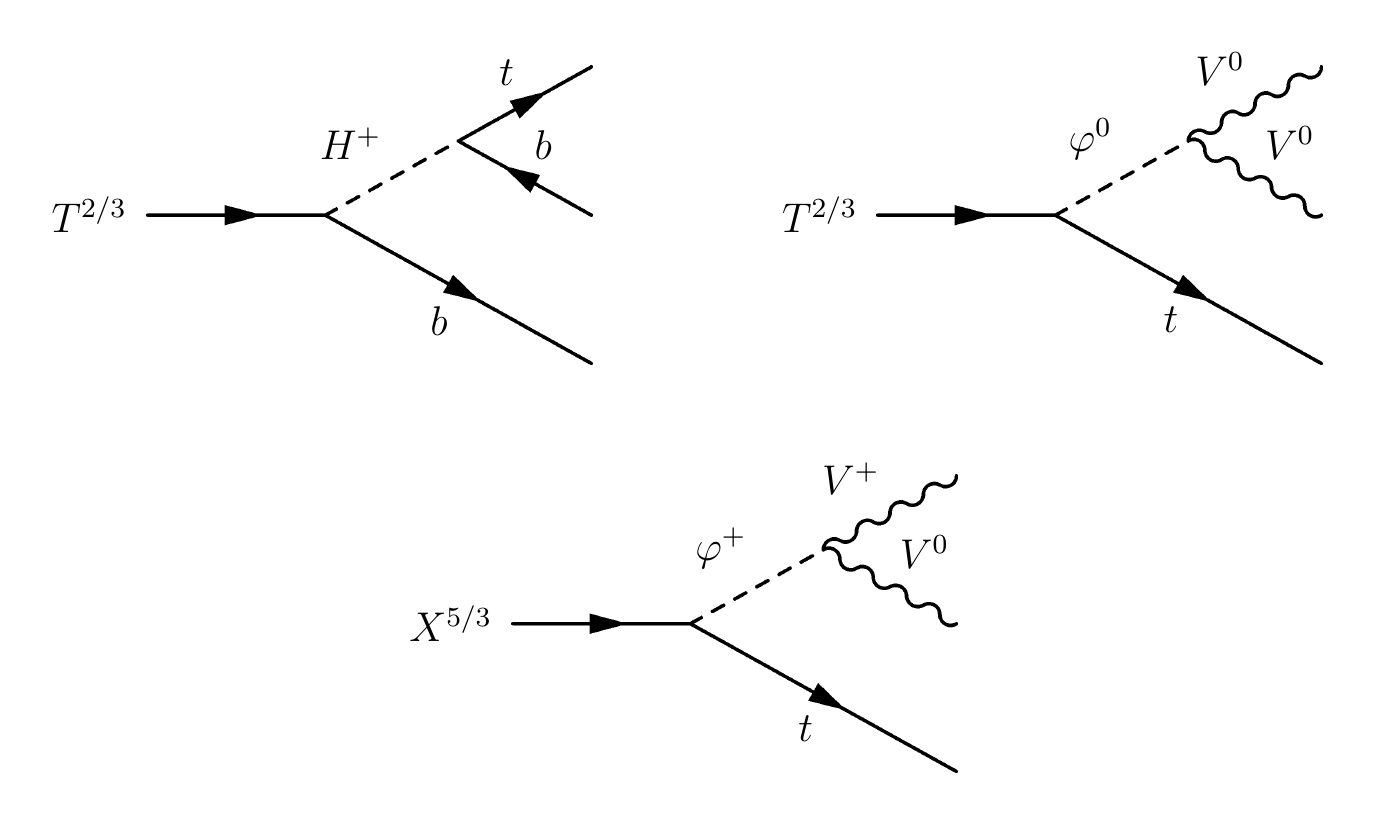}
\caption{Three examples of potentially interesting and relevant exotic top partner decays.  $T^{2/3}$ denotes a charge-$2/3$ top partner and $X^{5/3}$ a charge-$5/3$ top partner.  $\varphi$ represents a generic quarticon and $V$ a light Higgs or electroweak gauge boson.  The upper-left diagram yields the same final state as the decay $T \rightarrow t h \rightarrow t b b$, but with different $b$-jet kinematics.  The upper-right and lower diagrams yield novel, striking top partner decay topologies.  In particular, the lower diagram yields an extra electroweak boson relative to the usually-assumed $X^{5/3} \rightarrow t W^+$ decay.}
\label{fig:Tdecaydiagrams}
\end{figure}

Three particularly interesting decay patterns are depicted in \Fig{fig:Tdecaydiagrams}.  For instance, the decays $T^{2/3} \rightarrow b H^+ \rightarrow b t \bar{b}$ and $T^{2/3} \rightarrow t A^0 \rightarrow t b \bar{b}$ yield the same final state as $T^{2/3} \rightarrow t h, t Z \rightarrow t b \bar{b}$.  In principle, this could make searches for these decays challenging as they would suffer not only from SM backgrounds (particularly $pp \rightarrow tt+$jets and $pp \rightarrow ttbb$), but also from backgrounds due to (potentially dominant) vanilla top partner decays.  However, as the $b\bar{b}$ pair produced in the exotic decay will not necessarily exhibit $m_{bb} \approx m_h \text{ or } m_Z$, backgrounds can be suppressed by exploiting the high-$b$ multiplicities and large $b \bar{b}$ invariant masses of the signal process.  While detecting second Higgs doublet states in top partner decays is liable to be difficult, it may well be possible at the LHC with $\sqrt{s} = 14 \text{ TeV}$ and integrated luminosity of 300~fb$^{-1}$ \cite{Kearney:2013oia}.

Cascade decays involving quarticons can be particularly dramatic.  Since quarticons decay to pairs of electroweak bosons, this leads to striking signatures involving large numbers of jets (including $b$-jets), multiple leptons (including like-sign dileptons),  and missing energy.  For instance, the decay $T^{2/3} \rightarrow t \varphi^0 \rightarrow t h h$ would yield $5bW$ and $3b3W$ (with one $W$ off-shell) final states.  Also noteworthy are decays such as $X^{5/3} \rightarrow t \varphi^+ \rightarrow t W^+ Z$, which yields an extra electroweak boson relative to the usually-considered $X^{5/3} \rightarrow t W^+$ decay.  Of course, because top partners are pair-produced in QCD processes, there are significant combinatoric challenges to unambiguously reconstructing the cascade decay.

\section{A Concrete Example: $\SO(10)/\SO(5)^2$}
\label{sec:so10modso5}

We now present a motivating example model that exhibits all of the features discussed in the previous sections:  a slew of exotic top partners, numerous additional PNGBs, and the associated interesting phenomenology.  This model fills a gap present in the existing LH literature by showing how a ``Littlest Higgs'' can simultaneously have two Higgs doublets and a custodial symmetry.  The minimal model that satisfies these criteria involves the coset space $\SO(10)/\SO(5)^2$, and $\SO(10)$-invariant Yukawa couplings in the top sector yields exotic top partner phenomenology.

\subsection{Method of the Missing Box}
\label{subsec:LHpattern}

Consider the following algorithm for $G/H$ Littlest Higgs model building, a version of which was originally presented in \Ref{ArkaniHamed:2002qy}.  To implement collective breaking in the gauge sector (and avoid quadratically-divergent contributions to $m_h^2$ from electroweak gauge bosons at one loop), the global symmetry $G$ must contain two copies of a weakly-gauged subgroup $W_1, W_2 = \SU(2)$, whose diagonal subgroup is $W_V = \SU(2)_L \subset H$.  Each $W_i$ must commute with a different subgroup of $G$  (denoted $X_{i}$) acting non-linearly on the Higgs, such that only when both $W_i$ are gauged are all of the global symmetries acting on the Higgs broken.  Thus, $G$ contains two different but overlapping product subgroups: $W_1 \times X_1$ and $W_2 \times X_2$.  In fact, to ensure that the Higgs has the correct electroweak quantum numbers under $W_V = \SU(2)_L$, each $X_i$ must contain the other  $W_j$ ($j \neq i$) with some generators transforming as doublets of $W_j$.  Since both gauge couplings $g_1$ and $g_2$ must enter in any process that radiatively generates the Higgs potential, the one-loop divergences from gauge interactions are at worst logarithmic.  

\begin{table}
\centering
\renewcommand{\arraystretch}{1.5}
\begin{tabular}{c | c c | c c |}
\multicolumn{1}{c}{} & \multicolumn{2}{c}{1HDM} & \multicolumn{2}{c}{2HDM} \\ \cline{2-5}
\multicolumn{1}{r |}{\multirow{2}{*}{Non-custodial, $W_i = \SU(2)$}} & $X_i = \SU(3)$ & \multicolumn{1}{ c |}{\multirow{2}{*}{\cite{ArkaniHamed:2002qy}}} & $X_i = \SU(4)$ & \multicolumn{1}{ c |}{\multirow{2}{*}{\cite{Low:2002ws}}} \\ & $G = \SU(5)$ & & $G = \SU(6)$ & \\
\hhline{~----}
\multicolumn{1}{r |}{\multirow{2}{*}{Custodial, $W_i = \SO(4)$}} & $X_i = \SO(5)$ & \multicolumn{1}{ c |}{\multirow{2}{*}{\cite{Chang:2003zn}}} & \cellcolor[gray]{0.9} $X_i = \SO(6)$ & \multicolumn{1}{ c |}{\multirow{2}{*}{ \cellcolor[gray]{0.9} }} \\ & $G = \SO(9)$ & & \cellcolor[gray]{0.9} $G = \SO(10)$ &  \cellcolor[gray]{0.9} \\ \hhline{~----}

\end{tabular}
\caption{Possible simple group candidates for $G$ to produce a ``Littlest Higgs'' model that is (a) either a one (1HDM) or two (2HDM) Higgs doublet model and (b) either not custodially symmetric ($W_i = \SU(2)$) or custodially symmetric ($W_i = \SO(4)$).  The shaded ``missing box'' is the new $\SO(10)/\SO(5)^2$ model that we construct in this paper.}
\label{tab:boxmodel}
\end{table}

Following this reasoning, \Ref{ArkaniHamed:2002qy} concluded that constructing a theory with a single Higgs doublet required $X_i = \SU(3)$, making $G = \SU(5)$ the obvious candidate, leading to the $\SU(5)/\SO(5)$ Littlest Higgs (see \Tab{tab:boxmodel}).  Alternatively, producing a theory with two Higgs doublets (i.e.~requiring $X_i$ to contain two sets of generators transforming as doublets under $W_j$) leads to $X_i = \SU(4)$, making $G = \SU(6)$ the obvious candidate.  This is the symmetry group on which the $\SU(6)/\Sp(6)$ model of \Ref{Low:2002ws} is based.  However, these two models do not exhibit custodial symmetry $\SU(2)_L \times \SU(2)_R \subset H$, generically leading to tension with precision electroweak measurements.  Ameliorating this tension requires either the introduction of a parity to forbid custodial symmetry violating operators in the case of \Refs{ArkaniHamed:2002qy,Cheng:2003ju,Cheng:2004yc}, or that the parameters of the theory reside in an approximately custodially-symmetric region of parameter space in the case of \Refs{Low:2002ws,Gregoire:2003kr}.  

If we do not wish $G$-breaking to lead to custodial symmetry violation, then $H$ must contain an $\SU(2)_L \times \SU(2)_R \cong \SO(4)$ subgroup under which the Higgs field transforms as a $\mathbf{4}$.  By promoting $W_1, W_2 = \SO(4)$ (which are not necessarily entirely gauged but which do contain gauged $\SU(2)$ subgroups), we can construct ``Littlest Higgs'' models with custodial symmetry.  Requiring a single Higgs implies $X_i = \SO(5)$, and the obvious candidate is $G = \SO(9)$.  This case was considered in the $\SO(9)/(\SO(5) \times \SO(4))$ model in \Ref{Chang:2003zn}.  Unfortunately, this model is not viable because it has a ``dangerous singlet'' \cite{Schmaltz:2008vd} (see footnote~\ref{footnote:whyDangerous}).

Now looking at \Tab{tab:boxmodel}, we see there is a ``missing box'', and we can complete the pattern of Littlest Higgs models by considering a coset space that allows for both two Higgs doublets and a custodial symmetry.  By extending $X_i = \SO(6)$ to permit a second Higgs field, we arrive at the previously unconsidered candidate for a ``Littlest Higgs'' model with $G = \SO(10)$.   As further motivation to consider this model, dangerous singlets can be avoided in two Higgs doublet models, because a parity can be invoked to prevent a singlet tadpole (see footnote~\ref{footnote:whytwoHiggs}).  This is our starting point for a new $\SO(10)/\SO(5)^2$ construction.

\subsection{The $\SO(10)/\SO(5)^2$ Littlest Higgs}
\label{subsec:105model}

Motivated by the aforementioned pattern, we briefly outline the  ``missing Littlest Higgs'' based on the coset space $\SO(10)/\SO(5)^2$.

Consider a field $\Sigma$ that transforms as
\begin{equation}
\Sigma \rightarrow V \Sigma V^T
\end{equation}
under $G = \SO(10)$.  $\SO(10)$ is broken to $\SO(5)^2$ by the vev for $\Sigma$,
\begin{equation}
\setlength\arraycolsep{10pt}
\vev{\Sigma} = \left(\begin{array}{c c c c} & & & \mathbf{1}_4 \\ & 0 & 1 & \\ & 1 & 0 & \\ \mathbf{1}_4 & & & \end{array}\right),
\end{equation}
where $\mathbf{1}_4$ denotes the $4 \times 4$ identity matrix.  In the language of \Sec{subsec:LHpattern}, $W_1$ and $W_2$ are the $\SO(4)$'s living in the upper-left and lower-right blocks of $\SO(10)$, and $H = \SO(5)^2$ contains their diagonal subgroup $\SO(4)_V$ (which is identified with the custodial $\SU(2)_L \times \SU(2)_R$ symmetry of the SM).  The unbroken generators $T^a$ and the broken generators $X^a$ of $\SO(10)$ satisfy the relations
\begin{align}
T^a \vev{\Sigma} - \vev{\Sigma} T^a & = 0, \\
X^a \vev{\Sigma} + \vev{\Sigma} X^a & = 0.
\end{align}

From the broken generators, we can write a \nlsm\ field for the Goldstone multiplet parameterizing the coset space $\SO(10)/\SO(5)^2$ as
\begin{equation}
\Sigma = e^{2 i \Pi/f} \vev{\Sigma},
\end{equation}
 where
\begin{equation}
\setlength\arraycolsep{10pt}
i \Pi = i \pi^a X^a = \left(\begin{array}{c c c c} \omega_L + \eta_R & h_1 & h_2 & \phi +  \varphi^0 \mathbf{1}_4 \\ - h_1^T & 0 & \sigma & h_2^T \\ -h_2^T & - \sigma & 0 & h_1^T \\ - \phi - \varphi^0 \mathbf{1}_4 & -h_2 & - h_1 & -\omega_L - \eta_R\end{array}\right).
\end{equation}
Here, $(\omega_L + \eta_R)$ is an anti-symmetric $4 \times 4$ matrix, $\phi$ is a traceless symmetric $4 \times 4$ matrix, $h_1, h_2$ are 4 component (column) vectors, and $\sigma$ and $\varphi^0$ are real singlets.\footnote{For clarity, we have omitted factors necessary to canonically normalize the fields, since these factors are readily determined from the kinetic terms for $\Sigma$.}  As desired, $h_1$ and $h_2$ are the two Higgs doublets, each transforming as a $(\mathbf{2},\mathbf{2})$ of $\SO(4)_V \cong \SU(2)_L \times \SU(2)_R$.  The anti-symmetric matrix $\omega_L + \eta_R$ can be decomposed into two triplets transforming as $(\mathbf{3},\mathbf{1})$ ($\omega_L$) and $(\mathbf{1},\mathbf{3})$ ($\eta_R$).  The symmetric matrix $\phi$ transforms as a $\mathbf{9}$ of $\SO(4)_V$ or, equivalently, a $(\mathbf{3},\mathbf{3})$ of $\SU(2)_L \times \SU(2)_R$.  This accounts for the $45 - 2 \times 10 = 25$ PNGBs.

The fields $\phi$, $\varphi^0$, and $\sigma$ all transform appropriately under the shift symmetries protecting the Higgs multiplets to serve as quarticons (see \Sec{subsec:exoticpngbs}), but the two singlets $\sigma$ and $\varphi^0$ are potentially ``dangerous''.\footnote{The analog of $\varphi^0$ is indeed a dangerous singlet in the $\SO(9)/(\SO(5) \times \SO(4))$ model \cite{Chang:2003zn}.}  However, the chosen breaking pattern ensures that the $\SO(10)/\SO(5)^2$ Littlest Higgs admits a parity given by $\Sigma \rightarrow K \Sigma K$ with
\begin{equation}
\label{eq:Kparity}
\setlength\arraycolsep{10pt}
K = \left(\begin{array}{c c c c} \mathbf{1}_4 & & & \\ & 1 & & \\ & & -1 & \\ & & & -\mathbf{1}_4 \end{array}\right)
\end{equation}
under which the PNGB fields transform as
\begin{align*}
h_1 & \rightarrow h_1, & \omega_L & \rightarrow \omega_L, & \eta_R & \rightarrow \eta_R, \\ 
h_2 & \rightarrow -h_2, & \sigma & \rightarrow - \sigma, & \phi & \rightarrow - \phi, & \varphi^0 & \rightarrow - \varphi^0.
\end{align*}
This parity prevents $\sigma$ and $\varphi^0$ from being dangerous, and is the extension of the parity presented in \Ref{Schmaltz:2010ac} to a ``littlest'' structure.  By expanding the $G/H$ coset space to contain two Higgs bosons, we have constructed a ``Littlest Higgs'' model which exhibits custodial symmetry while avoiding dangerous singlets.

The SM gauge group is identified with the subgroup $\SU(2)_L \times \U(1)_R \subset \SO(4)_V$, which determines the electroweak quantum numbers of the PNGBs---the various PNGBs, their quantum numbers under $\SU(2)_L \times \SU(2)_R$ and their components are given in \Tab{tab:so10pngbs}.  Here, we abstain from additional detailed model building in the gauge and Higgs sectors as the main ingredients needed to complete the model mimic those exhibited by other LH models. We do, however, provide a sketch of the recipe in \App{app:so10gaugeandhiggs}.  Depending on the details of the gauge sector, and whether or not modular breaking is implemented, the components of $\omega_L$ and $\eta_R$ can either be eaten or remain in the spectrum as uneaten PNGBs.\footnote{While the uneaten modes will technically be a linear combination of $\omega_L$ and $\eta_R$ and components of another \nlsm\ field $\Delta$, for $V \gg f$ the uneaten modes will be predominantly $\omega_L$ and $\eta_R$.}

\begin{table}
\centering
\renewcommand{\arraystretch}{1.2}
\begin{tabular}{|| c | c || c | c || c ||} \hline
Type of PNGB & Field & $\SU(2)_L$ & $\SU(2)_R$ & Components \\ \hline \hline
\multirow{2}{*}{Higgs Doublets} & $h_1$ & $\mathbf{2}$ & $\mathbf{2}$ & \multirow{2}{*}{$h, H^0, A^0, Z_L, H^\pm, W_L^\pm$} \\
& $h_2$ & $\mathbf{2}$ & $\mathbf{2}$ & \\ \hline
\multirow{3}{*}{Quarticons} & $\phi$ & $\mathbf{3}$ & $\mathbf{3}$ & $\left\{\begin{array}{l}
\mathbb{C} \text{ triplet, } T^3_R = \pm 1: \phi^0_{\mathbb{C}}, \phi_{\mathbb{C}}^\pm, \phi_{\mathbb{C}}^{\pm \pm} \\
\mathbb{R} \text{ triplet, } T^3_R = 0: \phi_{\mathbb{R}}^0, \phi_{\mathbb{R}}^\pm \end{array}\right.$ \\
& $\varphi^0$ & $\mathbf{1}$ & $\mathbf{1}$ & $\varphi^0$ \\
& $\sigma$ & $\mathbf{1}$ & $\mathbf{1}$ & $\sigma$ \\ \hline
Potentially & $\omega_L$ & $\mathbf{3}$ & $\mathbf{1}$ & $\omega_L^0, \omega_L^{\pm}$ \\
Uneaten PNGBs & $\eta_R$ & $\mathbf{1}$ & $\mathbf{3}$ & $\eta_R^0, \eta_R^\pm$ \\ \hline
\end{tabular}
\caption{\label{tab:so10pngbs}Table of PNGBs resulting from breaking of $\SO(10) \rightarrow \SO(5)^2$.  The SM gauge group is identified with $\SU(2)_L \times \U(1)_R$, where $\U(1)_R \subset \SU(2)_R$ is  associated with the $T^3_R$ generator of $\SU(2)_R$.  $W_L^\pm$ and $Z_L$ denote the longitudinal components of the $W^\pm$ and $Z$ bosons respectively, corresponding to the eaten components of the Higgs multiplets.  Depending on which subgroups of $G$ are gauged and whether or not modular breaking is implemented, components of $\omega_L$ or $\eta_R$ can either be eaten or remain in the spectrum as uneaten PNGBs.  Neutral scalars are real except for $\phi^0_{\mathbb{C}}$.}
\end{table}

\subsection{Top Sector}
\label{subsec:so10topsector}

Following \Sec{subsec:exotictops}, we now construct a top sector for the $\SO(10)/\SO(5)^2$ model consisting of the interactions given in \Eqs{eq:GinvYukawa}{eq:softGbreaking},
\begin{equation}
\label{eq:so10topsector}
\mathcal{L}_{\text{top}} = - y_1 f \mathcal{Q}^{T} \Sigma \mathcal{Q}^c - y_2 f U_5 u^c - y_3 f q Q^c_{4,+} + \text{h.c.},
\end{equation}
where $\mathcal{Q}^{T} = (X^{T}_4, U_6, U_5, Q^{T}_4)$ and $\mathcal{Q}^{c \; T} = (Q^{c \; T}_4, U^c_5, U^c_6, X^{c \; T}_4)$ transform as $\mathbf{10}$'s of $\SO(10)$.  $Q^{(c)}_4$ and $X^{(c)}_4$ are 4-component vectors, each transforming as a $\mathbf{4}$ under $\SO(4)_V$ (i.e.~a $(\mathbf{2},\mathbf{2})$ under $\SU(2)_L \times \SU(2)_R$).  We use $Q^c_{4,\pm}$ to denote the $\SU(2)_L$ doublet with $T_R^3 = \pm \frac{1}{2}$.  $U^{(c)}_5$ and $U^{(c)}_6$ are custodial singlets.
As in \Sec{subsec:exotictops}, $q$ transforms as a $(\mathbf{2},\mathbf{1})$ under $\SU(2)_L \times \SU(2)_R$ and $u^c$ is a custodial singlet.  $\SU(2)_L$ doublets are contracted with $\epsilon_{ij}$.

In order to produce the correct hypercharge assignments for the quark multiplets, we must gauge a linear combination of $T_R^3$ and an additional global $\U(1)_X$ acting on the fermions
\begin{equation}
T_Y \equiv T_R^3 + T_X.
\end{equation}
Under this $\U(1)_X$, fields without a superscript $c$ have charge $+\frac{2}{3}$ and those with a superscript $c$ have charge $-\frac{2}{3}$.

Setting $\Sigma$ equal to its vev gives the leading mass terms for the $\mathcal{Q}^c$ and $\mathcal{Q}$ components,
\begin{equation}
\mathcal{Q}^{T} \vev{\Sigma} \mathcal{Q}^c = Q_{4,-} Q^c_{4,+} - Q_{4,+} Q^c_{4,-} + X_{4,-} X^c_{4,+} - X_{4,+} X^c_{4,-} + U_5 U^c_5 + U_6 U^c_6.
\end{equation}
We thus  expect a series of exotic fields at mass $y_{1} f$ (more on these below).  As usual, we define fields as in \Eq{eq:topfieldredef} (for simplicity, we shall assume the $y_i$ are real and positive),
\begin{align}
\label{eq:topfieldredefso10}
Q_3 & \equiv \frac{y_1 Q_{4,-} + y_3 q}{\sqrt{y_1^2 + y_3^2}}, &
q_3 & \equiv \frac{y_3 Q_{4,-} - y_1 q}{\sqrt{y_1^2 + y_3^2}}, &
T^c & \equiv \frac{y_1 U^c_5 + y_2 u^c}{\sqrt{y_1^2 + y_2^2}}, & 
t^c & \equiv \frac{y_2 U^c_5 - y_1 u^c}{\sqrt{y_1^2 + y_2^2}},
\end{align}
allowing us to identify mass eigenstates prior to electroweak symmetry breaking (i.e.~to leading order in $v/f$).  The fields $q_3$ and $t^c$ have the correct quantum numbers to be identified as SM top fields to leading order---they are an electroweak doublet with $T_Y = +\frac{1}{6}$ and a singlet with $T_Y = -\frac{2}{3}$ respectively.
Consequently, the cancellons will be the vector-like doublet consisting of $Q_3$ married to $Q^c_{4,+}$ and the vector-like singlet consisting of $U_5$ married to $T^c$.
For clarity, we relabel the other (non-cancellon) components of $Q_4$ (needed to form a complete multiplet of $\SU(2)_L \times \SU(2)_R$) as
\begin{equation}
Q_{4,+} \equiv Y_{4,+}, \qquad Q^c_{4,-} \equiv Y^c_{4,-},
\end{equation}
to distinguish them from the cancellon components of $Q_4$. The couplings in $\mathcal{L}_{\text{top}}$ break all of the shift symmetries acting on the first Higgs doublet $h_1$, so they will radiatively generate a negative $m_{h_1}^2$, but as long as other positive contributions dominate (e.g.~from the gauge sector, see \App{app:so10gaugeandhiggs}) the vacuum will be stable.

\begin{table}
\centering
\renewcommand{\arraystretch}{1.1}
\begin{tabular}{|| c | c | c | c | c | l l l l ||}
\hline
Name & Fields & Charge & Mass & Cancellon? & \multicolumn{4}{c ||}{Decays} \\ \hline \hline
\multirow{2}{*}{$X_1$} &\multirow{2}{*}{$X_{4,+}/X^c_{4,-}$} & $5/3$ & $y_1 f$ &  & $\phi_{\mathbb{C}}^{++} b$, & $\phi_{\mathbb{C}}^+ t$, & $V^+ t$ & \\
&& $2/3$ & $y_1 f$ &   & $\phi_{\mathbb{C}}^+ b$, & $\phi_{\mathbb{C}}^0 t$, & $V^0 t$ & \\
\hline
\multirow{2}{*}{$X_2$} &\multirow{2}{*}{$X_{4,-}/X^c_{4,+}$} & $2/3$ & $y_1 f$ &   & $\phi_{\mathbb{R}}^+ b$, & $\phi_{\mathbb{R}}^0 t$, & $\varphi^0 t$, & $V^0 t$ \\
&& $1/3$ & $y_1 f$ &   & $\phi_{\mathbb{R}}^- t$, & $\phi_{\mathbb{R}}^0 b$, & $\varphi^0 b$, & $V^- t$ \\
\hline
$U_6$ &$U_6/U^c_6$ & $2/3$ & $y_1 f$ &   & $\sigma t$, & $V^+ b$, & $V^0 t$ & \\
\hline
\multirow{2}{*}{$Y$} &\multirow{2}{*}{$Y_{4,+}/Y^c_{4,-}$}   & $5/3$ & $y_1 f$ &   & $\eta_R^+ t$, & $V^+ t$ & & \\
& & $2/3$ & $y_1 f$ &  & $\eta_R^+ b$, & $V^0 t$ & & \\
\hline
$T$ &$U_5/T^c$ & $2/3$ & $\sqrt{y_1^2 + y_2^2} f$ & \checkmark  & $V^+ b$, & $V^0 t$ & & \\
\hline
\multirow{2}{*}{$Q_3$} &\multirow{2}{*}{$Q_3/Q^c_{4,+}$}  & $2/3$ & $\sqrt{y_1^2 + y_3^2} f$ & \checkmark & $\omega_L^+ b,$ & $\omega_L^0 t$, & $\eta_R^0 t$, & $V^0 t$ \\
&& $1/3$ & $\sqrt{y_1^2 + y_3^2} f$ & \checkmark  & $\omega_L^- t,$ & $\omega_L^0 b$, & $\eta_R^0 b$, & $V^- t$ \\
\hline
\end{tabular}
\caption{\label{tab:so10toppartners} Charges, masses (to leading order in $v/f$), and most relevant decays for the vector-like top partners (cancellons and exotics) with the top sector in \Eq{eq:so10topsector}.  $V$ denotes (longitudinal) electroweak bosons ($h, Z_L$, or $W_L^\pm$), but also second Higgs doublet states $H^0, A^0$, or $H^\pm$.  As all top partners couple to Higgs multiplets, the usual top partner decays will still occur.  However, different top partners couple to different Higgs doublets (see \App{app:so10toppartnerinteractions} for details), so the significance of the usual decays will depend on, e.g., $\tan \beta$.  Consequently, only the singlet cancellon $T$ is expected to exhibit the usual top partner decay pattern, $T \rightarrow b W^+, t h, t Z$ in an approximate $2:1:1$ ratio.}
\end{table}

With a top sector of this form, the spectrum contains a large multiplicity of new states. The vector-like singlet ($T$) and doublet ($Q_3$) cancellons are accompanied by seven vector-like non-cancellons, arranged into three doublets ($X_1, X_2$, and $Y$) and a singlet ($U_6$).  Two of these doublets ($X_1$ and $Y$) are necessary to complete four-plets of the custodial $\SO(4)_V$, and so necessarily contain exotic charge-$5/3$ quarks.  The constituent fields, charges, and masses (prior to electroweak symmetry breaking) for the various top partners are summarized in \Tab{tab:so10toppartners}.  Also shown are the likely dominant decay modes for each top partner.  These can be determined by examining the marginal operators present in \Eq{eq:so10topsector} and redefining fields as in \Eq{eq:topfieldredefso10}, with details given in \App{app:so10toppartnerinteractions}.  The additional PNGBs decay as outlined in \Tab{tab:decaytopologies}.

The decay modes listed in \Tab{tab:so10toppartners} all arise from renormalizable couplings and are leading order in $v/f$, such that exotic decays are likely significant.\footnote{Exotic decays can also arise from higher-dimension operators, but would be suppressed by phase space or inverse powers of $f$.  In addition, after electroweak symmetry breaking, all of the charge-$2/3$ quarks will mix---in principle, this also permits exotic top partner decays, although suppressed by factors of $v/f$.}
Exact branching ratios will depend on the specific values of $y_i$, $\tan \beta$, mixing between Higgs states, and phase space suppression (as the additional PNGBs and second Higgs doublet states are expected to be somewhat heavy).
All top partners exhibit renormalizable couplings to SM fields ($q_3$ or $t^c$) and Higgs doublets, so can exhibit the usual decays to third generation quarks and (longitudinal) electroweak bosons ($h, Z_L$ or $W_L^\pm$).  However, as different top partners couple to different Higgs multiplets (see \App{app:so10toppartnerinteractions}), if some decay dominantly to SM bosons then others may well decay more frequently to second Higgs doublet states (depending on the details of the Higgs sector).
In fact, it is reasonable to expect that only the singlet cancellon $T$ will exhibit the usual decay pattern, namely $T \rightarrow b W^+, th, tZ$ in an approximate $2:1:1$ ratio.

Furthermore, even if cancellons decay predominantly via $T \rightarrow b W^+, th, tZ$, they can exhibit sub-dominant decays to extended Higgs sector states  \cite{Kearney:2013oia}.  Both singlet and doublet cancellons can decay to second Higgs doublet states, and the doublet cancellon can also decay to uneaten PNGBs if they are present.  This motivates searches for such exotic and sub-dominant decays.  As they are heavier than the non-cancellons, cancellons could conceivably undergo cascade decays to other top partner states, although such decays are expected to be highly phase-space suppressed.

For the non-cancellons, $G$-invariance and the numerous additional PNGBs give rise to a wide variety of possible exotic decays outlined in \Tab{tab:so10toppartners}.  Consequently, this model combines all of the features anticipated in \Secs{subsec:exotictops}{subsec:exoticpngbs}, giving rise to the exotic phenomenology outlined in \Sec{sec:exoticpheno} where top partner pair production yields a final state with a high multiplicity of third-generation quarks and electroweak bosons.

\section{Conclusions}
\label{sec:conclusions}

In this paper, we have highlighted theoretical arguments to suggest that realistic LH models will contain (1) top partners beyond the minimal set required to cancel quadratic divergences in the Higgs sector and (2) PNGBs beyond a single Higgs doublet.  The existence of additional PNGBs is mandatory.  At minimum, LH models must contain quarticons to generate a collective Higgs quartic coupling and maintain the parametric hierarchy $v \ll f$.  A second Higgs doublet and uneaten PNGBs are also well-motivated, given the model building goals of preserving custodial $\SU(2)$ and lifting gauge partners through modular breaking.  The existence of additional top partners, while strictly speaking not necessary, follows from a very simple and motivating assumption that the strong dynamics is $G$-preserving up to operators that mix composite and elementary states.  Thus, it is reasonable to expect the strong dynamics to generate $G$-invariant couplings, in which case some fermions must transform in complete multiplets of $G$.  As LH theories frequently feature enlarged symmetry groups $G$, these long top multiplets will generally contain top partners beyond the minimal set.  

In addition to providing exotic top partners, the large $G$ symmetry could have interesting implications for phenomenology.  With top partners and PNGBs all transforming under $G$, $G$-invariance necessarily implies the existence of couplings which could allow exotic top partner decays beyond the standard $T \rightarrow bW^+$, $th$, and $tZ$ modes usually considered.  Exotic decays involving the additional PNGBs can exhibit different event kinematics than standard decays, and may even produce extraordinary final states rich in $b$-jets and electroweak bosons.  In particular, events with many $b$-jets (but different kinematics from $T \rightarrow bW^+$, $th$, and $tZ$) may be indicative of a second Higgs doublet or uneaten PNGBs, since both types of additional PNGBs likely decay to third-generation quarks.  A corollary is that experimental searches should avoid imposing requirements like $m_{bb} = m_{h} \text{ or } m_Z$ to the extent possible.  Top partner decays may be the best way to discover new Higgs multiplets or more exotic PNGBs, as electroweak production of such states is likely to be limited at the LHC.  Consequently, searches for exotic top partner decays may prove vital to our understanding of the strong dynamics that yield a PNGB Higgs boson.

\acknowledgments
We would like to acknowledge useful conversations with Spencer Chang and Martin Schmaltz. The work of AP and JK  is supported in part by the NSF CAREER grant number NSF-PHY-0743315.  The work of AP is also supported in part  by DoE grant number DE-SC0007859.  The work of JT is supported by the U.S. DoE under cooperative research agreement DE-FG02-05ER-41360 and under the DoE Early Career research program DE-FG02-11ER-41741.

\appendix

\section{Details of $\SO(10)/\SO(5)^2$ Construction}
\label{app:so10gaugeandhiggs}

In this appendix, we give further details on the $\SO(10)/\SO(5)^2$ Littlest Higgs from \Sec{subsec:105model}.  The SM gauge group is obtained by gauging subgroups $[\SU(2) \times \U(1)]_i \subset W_i$ such that the diagonal $\SU(2)_L \times \U(1)_R \subset \SO(4)_V$ is left unbroken by $\vev{\Sigma}$ and can be identified as the SM electroweak group.  Quadratically-divergent radiative corrections due to gauge interactions will generate operators as in \Eq{eq:quarticonL} with $h^2$ replaced by $h_1 h_2^T$ ($h_1^T h_2$) and with $\varphi$ replaced by $\phi$ ($\varphi^0$).  However, the corresponding mass terms for $\phi$ and $\varphi^0$ will have opposite signs such that one is necessarily tachyonic \cite{Cheung:2006ij}.  This is a manifestation of a vacuum alignment problem---if $\vev{\Sigma}$ were proportional to the identity, then both $W_1$ and $W_2$ (and hence the gauge groups) would be unbroken.  The tachyonic direction can be lifted by inserting ``plaquette operators'' by hand in such a way as to give additional (positive) contributions to $\lambda_\pm$ (ensuring that the operators maintain at least one of the shift symmetries protecting the Higgs, as in \Refs{ArkaniHamed:2001nc,Chang:2003zn,Low:2002ws}).\footnote{An alternative solution would be to go to still larger $G$ (such as $\SU(9)$ \cite{Chang:2003zn}) with $\Sigma$ still transforming as $\Sigma \rightarrow V \Sigma V^T$---such a model would not exhibit a vacuum alignment problem.}

Once the tachyonic directions have been stabilized, $\phi$ and $\varphi^0$ will both act as quarticons---integrating them out will collectively generate a Higgs quartic.  An additional contribution to the quartic can be introduced via operators of the form
\begin{equation}
\lambda_{55} f^4 (\Sigma_{55})^2 + \lambda_{66} f^4 (\Sigma_{66})^2,
\end{equation}
where $\Sigma_{ij}$ represents the $(i,j)$ entry in $\Sigma$.  These operators will produce terms as in \Eq{eq:quarticonL} with $\varphi \rightarrow \sigma$ and $h^2 \rightarrow h_1^T h_2$.

Such terms have the added advantage of giving a mass to $\sigma$, which (being a singlet) does not acquire a mass due to gauge interactions.  Gauge interactions will, however, radiatively generate logarithmically divergent masses for $h_1$ and $h_2$ of the form
\begin{equation}
m_1^2 h_1^T h_1 + m_2^2 h_2^T h_2
\end{equation}
at one-loop, as gauging subgroups of $W_1$ and $W_2$ collectively breaks all of the shift symmetries protecting the Higgs multiplets.  These masses and the collective quartic will also receive radiative corrections from the top sector.  

To destabilize the origin of field space and trigger electroweak symmetry breaking, it is necessary to induce a $m_{12}^2 h_1^T h_2$ mass term.  Such a term will necessarily not be generated radiatively (and must be inserted by hand) as all interactions are designed to respect the parity $\Sigma \rightarrow K \Sigma K$ from \Eq{eq:Kparity} in order to avoid dangerous singlets, whereas this $B \mu$-type term explicitly violates this parity.  However, as long as the parity is broken ``softly,'' the radiative contributions to the Higgs potential are suppressed.   This can be accomplished by the inclusion of operators such as
\begin{equation}
m_{55}^2 f^2 \Sigma_{55} + m_{66}^2 f^2 \Sigma_{66},
\end{equation}
which includes a $B \mu$-type term and a $\sigma$ tadpole.
So, while explicit $G$-breaking terms are needed to yield a phenomenologically realistic model (as in other LH constructions), these terms can be included in such a way that does not reintroduce one-loop quadratically-divergent contributions to the Higgs mass terms, maintaining the hierarchy $v \ll \Lambda \simeq 4 \pi f$.\footnote{In the case of modular breaking, mass terms may also be required for the uneaten components of $\omega_L$ and $\eta_R$---again, such masses can be generated radiatively or inserted by hand via plaquette operators \cite{Schmaltz:2010ac}.}

Finally, it is worth noting that the electroweak triplets $\phi$ can acquire a vev.\footnote{In the case of modular breaking, $\omega_L$ can also get a vev.  However, this will be suppressed as $\omega_L$ does not have tree-level couplings to the Higgs.}  Usually, triplet vevs are required to be significantly smaller than the Higgs vev by precision electroweak measurements.  Fortunately, as in \Ref{Chang:2003zn}, the approximate custodial symmetry ensures that custodial $\SU(2)$ violation from the triplet vevs is limited to be small, consistent with precision constraints.  However, direct measurements of Higgs couplings to gauge bosons may soon place additional stringent constraints on electroweak triplet vevs, even ones that preserve custodial $\SU(2)$.

\section{Top Partner Interactions in the $\SO(10)/\SO(5)^2$ Model}
\label{app:so10toppartnerinteractions}

In this appendix, we expand \Eq{eq:so10topsector} and redefine fields according to \Eq{eq:topfieldredefso10} in order to determine the most relevant decays of top partners to SM third-generation quarks ($q_3$ and $t^c$) and gauge bosons or PNGBs.  As in the text, we have not canonically normalized the PNGB fields.

First consider the $X$ fields.  To leading order, there are two vector-like doublets: $X_1$, consisting of $X_{4,+}$ married to $X^c_{4,-}$, and $X_2$, consisting of $X_{4,-}$ married to $X^c_{4,+}$.
Dimension four operators in $\mathcal{L}_{\text{top}}$ coupling these fields to the SM top fields include
\begin{equation}
\mathcal{L}_{\text{top}, X}^{(4)} \supset y_1 (Q_{4,-} \varphi^0 X^c_{4,+} + Q_{4,-} \phi_0 X^c_{4,+} + Q_{4,-} \phi_+ X^c_{4,-} - X_{4,-} H_2 U^c_5 + X_{4,+} \widetilde{H}_2 U^c_5) + \text{h.c.},
\end{equation}
where $\phi_\pm, \phi_0$ denote the three $\SU(2)_L$ triplets.  The subscripts on the $\phi$ label $T_R^3 = \pm 1, 0$, and $H_2$ denotes $h_2$ written as an $\SU(2)_L$ doublet with $T_R^3 = +\frac{1}{2}$ while $\widetilde{H}_2 \equiv i \sigma_2 H_2$ is $h_2$ written as an $\SU(2)_L$ doublet with $T_R^3 = -\frac{1}{2}$.  Using \Eq{eq:topfieldredefso10}, these interactions can be rewritten as
\begin{equation}
\mathcal{L}_{\text{top}, X}^{(4)} \supset \frac{y_1 y_3}{\sqrt{y_1^2 + y_3^2}} q_3 \left(\varphi^0 X^c_{4,+} + \phi_0 X^c_{4,+} +\phi_+ X^c_{4,-}\right) - \frac{y_1 y_2}{\sqrt{y_1^2 + y_2^2}} \left(X_{4,-} H_2 - X_{4,+} \widetilde{H}_2\right) t^c + \text{h.c.}
\end{equation}
The first pair of interactions will permit the decays of these non-cancellon top partners to quarticons and SM fields, including (using $X^q$ and $\phi^q$ to denote an $X$ quark or a component of $\phi$ with the superscript labeling the electric charge $q$)
\begin{align}
X_1^{+5/3} & \rightarrow \phi^+ t_3, \phi^{++} b_3, & X_{1,2}^{+2/3} & \rightarrow \phi^0 t_3, \phi^+ b_3, & X_2^{-1/3} & \rightarrow \phi^0 b_3, \phi^- t_3.
\end{align}
The second pair of interactions will permit top partner decays to $t^c$ in conjunction with the additional Higgs states present in a two Higgs doublet model, namely $H^0, A^0$ and $H^\pm$.  Of course, these interactions also permit ordinary decays to (longitudinal) electroweak bosons ($h$, $Z_L$, $W_L^\pm$).

We can perform a similar analysis for the $U_6$ and $Y$ fields.  $U_6$ consists of the vector-like singlet pair $U_6$ and $U^c_6$, which forms a Dirac quark of mass $y_1 f$ and charge $\pm \frac{2}{3}$.  We have
\begin{align}
\mathcal{L}_{\text{top}, U_6}^{(4)} & \supset y_1 \left(- U_6 \sigma U_5^c + Q_{4,-} H_2 U_6^c\right) + \text{h.c.}  \notag \\ 
& \supset - \frac{y_1 y_2}{\sqrt{y_1^2 + y_2^2}} U_6 \sigma t^c + \frac{y_1 y_3}{\sqrt{y_1^2 + y_3^2}} q_3 H_2 U_6^c + \text{h.c.},
\end{align}
permitting $U_6^{+2/3} \rightarrow \sigma t^c$ decays and decays to $t_3$ and $b_3$ in conjunction with both electroweak bosons and second Higgs doublet states.  Likewise, $Y$ consists of a vector-like doublet pair $Y_{4,+}$ and $Y^c_{4,-}$, also with Dirac mass $y_1 f$.  If the $\eta_R$ are (largely) uneaten due to the implementation of modular breaking, then $Y$ will have the interactions
\begin{align}
\mathcal{L}_{\text{top}, Y}^{(4)} & \supset y_1 \left(Q_{4,-} \eta_{R,+} Y^c_{4,-} - Y_{4,+} \widetilde{H}_1 U_5^c\right) + \text{h.c.} \notag \\
& \supset \frac{y_1 y_3}{\sqrt{y_1^2 + y_3^2}} q_3 \eta_{R,+} Y_{4,-}^c - \frac{y_1 y_2}{\sqrt{y_1^2 + y_2^2}} Y_{4,+} \widetilde{H}_1 t^c + \text{h.c.},
\end{align}
permitting decays such as $Y^{5/3} \rightarrow \eta_R^+ t_3$ and $Y^{2/3} \rightarrow\eta_R^+ b_3$.  These decays are in addition to decays to $t^c$ in conjunction with both electroweak bosons and second Higgs doublet states.  If the $\eta_R$ are eaten, then the dominant decay of $Y^{5/3}$ would likely be via $Y^{5/3} \rightarrow W^{+} t^c$ as usually expected.

Cancellon decays to third generation quarks can be determined by analyzing the interactions of $U_5/U^c_5$ and $Q_{4,-}/Q^c_{4,+}$ fields.  The singlet cancellon $T$ consists of $U_5$ married to $T^c$, and exhibits a renormalizable coupling
\begin{align}
\mathcal{L}^{(4)}_{\text{top}, U_5} & \supset y_1 Q_{4,-} H_1 U_5^c + \text{h.c.} \notag \\ 
& \supset \frac{y_1^2 y_3}{\sqrt{y_1^2 + y_2^2} \sqrt{y_1^2 + y_3^2}} q_3 H_1 T^c + \text{h.c.}
\end{align}
Thus, consistent with the usual expectation for top partner cancellons, the singlet cancellon will decay predominantly as $T^{2/3} \rightarrow W_L^+ b_3, h t_3, Z_L t_3$.  However, small $\tan \beta$ or non-decoupling between $H_1$ and $H_2$ could also lead to subdominant decays to second Higgs doublet states.  For the doublet cancellon $Q_3$, consisting of $Q_3$ married to $Q_{4,+}^c$, the situation is slightly more complicated if there are uneaten PNGBs in the spectrum.
The renormalizable Yukawa couplings are
\begin{align}
\mathcal{L}^{(4)}_{\text{top}, Q} & \supset y_1 \left(Q_{4,-} H_1 U_5^c + Q_{4,-} \omega_L Q_{4,+}^c + Q_{4,-} \eta_{R,0} Q_{4,+}^c\right) + \text{h.c.} \notag \\
& \supset \frac{y_1^2 y_2}{\sqrt{y_1^2 + y_2^2} \sqrt{y_1^2 + y_3^2}} Q_3 H_1 t^c + \frac{y_1 y_3}{\sqrt{y_1^2 + y_3^2}} q_3 \omega_L Q_{4,+}^c + \frac{y_1 y_3}{\sqrt{y_1^2 + y_3^2}} q_3 \eta_{R,0} Q_{4,+}^c + \text{h.c.}
\end{align}
The first term permits the usual top partner decays $Q_3^{2/3} \rightarrow h t^c, Z_L t^c$ and $Q_3^{-1/3} \rightarrow W_L^- t^c$ (and, as for $T$ above, potentially decays to second Higgs doublet states).  However, if there are uneaten PNGBs in the spectrum, the doublet cancellon can also exhibit exotic decays
\begin{align}
Q_3^{2/3} & \rightarrow \omega_L^0 t_3, \; \eta_R^0 t_3, \; \omega_L^+ b_3, & Q_3^{-1/3} & \rightarrow \omega_L^0 b_3, \; \eta_R^0 b_3, \; \omega_L^- t_3.
\end{align}

We have not shown interactions between cancellons and non-cancellons, which are indeed present.  Because cancellons are heavier than non-cancellons, such couplings permit cascade decays of cancellons to other exotic top partners, though the branching ratios for such cascade decays are likely small due to phase space suppression.  Finally, note that different top partners couple to different Higgs doublets: the cancellons and $Y$ fields couple to $H_1$, while the $X$ and $U_6$ fields couple to $H_2$.

\bibliography{LHBib}{}

\providecommand{\href}[2]{#2}\begingroup\raggedright\begin{thebibliography}{10}

\bibitem{Kaplan:1983fs}
D.~B. Kaplan and H.~Georgi, {\it {$SU(2) \times U(1)$ Breaking by Vacuum
  Misalignment}},  {\em Phys.Lett.} {\bf B136} (1984) 183.

\bibitem{Kaplan:1983sm}
D.~B. Kaplan, H.~Georgi, and S.~Dimopoulos, {\it {Composite Higgs Scalars}},
  {\em Phys.Lett.} {\bf B136} (1984) 187.

\bibitem{ArkaniHamed:2002qx}
N.~Arkani-Hamed, A.~Cohen, E.~Katz, A.~Nelson, T.~Gregoire, et~al., {\it {The
  Minimal moose for a little Higgs}},  {\em JHEP} {\bf 0208} (2002) 021,
  [\href{http://xxx.lanl.gov/abs/hep-ph/0206020}{{\tt hep-ph/0206020}}].

\bibitem{ArkaniHamed:2002qy}
N.~Arkani-Hamed, A.~Cohen, E.~Katz, and A.~Nelson, {\it {The Littlest Higgs}},
  {\em JHEP} {\bf 0207} (2002) 034,
  [\href{http://xxx.lanl.gov/abs/hep-ph/0206021}{{\tt hep-ph/0206021}}].

\bibitem{Schmaltz:2005ky}
M.~Schmaltz and D.~Tucker-Smith, {\it {Little Higgs review}},  {\em
  Ann.Rev.Nucl.Part.Sci.} {\bf 55} (2005) 229--270,
  [\href{http://xxx.lanl.gov/abs/hep-ph/0502182}{{\tt hep-ph/0502182}}].

\bibitem{Perelstein:2005ka}
M.~Perelstein, {\it {Little Higgs models and their phenomenology}},  {\em
  Prog.Part.Nucl.Phys.} {\bf 58} (2007) 247--291,
  [\href{http://xxx.lanl.gov/abs/hep-ph/0512128}{{\tt hep-ph/0512128}}].

\bibitem{Perelstein:2003wd}
M.~Perelstein, M.~E. Peskin, and A.~Pierce, {\it {Top quarks and electroweak
  symmetry breaking in little Higgs models}},  {\em Phys.Rev.} {\bf D69} (2004)
  075002, [\href{http://xxx.lanl.gov/abs/hep-ph/0310039}{{\tt
  hep-ph/0310039}}].

\bibitem{Han:2003wu}
T.~Han, H.~E. Logan, B.~McElrath, and L.-T. Wang, {\it {Phenomenology of the
  little Higgs model}},  {\em Phys.Rev.} {\bf D67} (2003) 095004,
  [\href{http://xxx.lanl.gov/abs/hep-ph/0301040}{{\tt hep-ph/0301040}}].

\bibitem{Meade:2006dw}
P.~Meade and M.~Reece, {\it {Top partners at the LHC: Spin and mass
  measurement}},  {\em Phys.Rev.} {\bf D74} (2006) 015010,
  [\href{http://xxx.lanl.gov/abs/hep-ph/0601124}{{\tt hep-ph/0601124}}].

\bibitem{DeSimone:2012fs}
A.~De~Simone, O.~Matsedonskyi, R.~Rattazzi, and A.~Wulzer, {\it {A First Top
  Partner's Hunter Guide}},  {\em JHEP} {\bf 1304} (2013) 004,
  [\href{http://xxx.lanl.gov/abs/1211.5663}{{\tt arXiv:1211.5663}}].

\bibitem{Buchkremer:2013bha}
M.~Buchkremer, G.~Cacciapaglia, A.~Deandrea, and L.~Panizzi, {\it {Model
  Independent Framework for Searches of Top Partners}},
  \href{http://xxx.lanl.gov/abs/1305.4172}{{\tt arXiv:1305.4172}}.

\bibitem{Berger:2012ec}
J.~Berger, J.~Hubisz, and M.~Perelstein, {\it {A Fermionic Top Partner:
  Naturalness and the LHC}},  {\em JHEP} {\bf 1207} (2012) 016,
  [\href{http://xxx.lanl.gov/abs/1205.0013}{{\tt arXiv:1205.0013}}].

\bibitem{ATLAS:2012qe}
{\bf ATLAS Collaboration} Collaboration, G.~Aad et~al., {\it {Search for pair
  production of heavy top-like quarks decaying to a high-pT $W$ boson and a $b$
  quark in the lepton plus jets final state at $\sqrt{s}=7$ TeV with the ATLAS
  detector}},  {\em Phys.Lett.} {\bf B718} (2013) 1284--1302,
  [\href{http://xxx.lanl.gov/abs/1210.5468}{{\tt arXiv:1210.5468}}].

\bibitem{Chatrchyan:2012vu}
{\bf CMS Collaboration} Collaboration, S.~Chatrchyan et~al., {\it {Search for
  pair produced fourth-generation up-type quarks in $pp$ collisions at
  $\sqrt{s}=7$ TeV with a lepton in the final state}},  {\em Phys.Lett.} {\bf
  B718} (2012) 307--328, [\href{http://xxx.lanl.gov/abs/1209.0471}{{\tt
  arXiv:1209.0471}}].

\bibitem{CMS:2012ab}
{\bf CMS Collaboration} Collaboration, S.~Chatrchyan et~al., {\it {Search for
  heavy, top-like quark pair production in the dilepton final state in $pp$
  collisions at $\sqrt{s} = 7$ TeV}},  {\em Phys.Lett.} {\bf B716} (2012)
  103--121, [\href{http://xxx.lanl.gov/abs/1203.5410}{{\tt arXiv:1203.5410}}].

\bibitem{Kaplan:2003uc}
D.~E. Kaplan and M.~Schmaltz, {\it {The Little Higgs from a simple group}},
  {\em JHEP} {\bf 0310} (2003) 039,
  [\href{http://xxx.lanl.gov/abs/hep-ph/0302049}{{\tt hep-ph/0302049}}].

\bibitem{Gregoire:2002ra}
T.~Gregoire and J.~G. Wacker, {\it {Mooses, topology and Higgs}},  {\em JHEP}
  {\bf 0208} (2002) 019, [\href{http://xxx.lanl.gov/abs/hep-ph/0206023}{{\tt
  hep-ph/0206023}}].

\bibitem{Katz:2003sn}
E.~Katz, J.-y. Lee, A.~E. Nelson, and D.~G. Walker, {\it {A Composite little
  Higgs model}},  {\em JHEP} {\bf 0510} (2005) 088,
  [\href{http://xxx.lanl.gov/abs/hep-ph/0312287}{{\tt hep-ph/0312287}}].

\bibitem{Thaler:2005en}
J.~Thaler and I.~Yavin, {\it {The Littlest Higgs in Anti-de Sitter space}},
  {\em JHEP} {\bf 0508} (2005) 022,
  [\href{http://xxx.lanl.gov/abs/hep-ph/0501036}{{\tt hep-ph/0501036}}].

\bibitem{Callan:1969sn}
J.~Callan, Curtis~G., S.~R. Coleman, J.~Wess, and B.~Zumino, {\it {Structure of
  phenomenological Lagrangians. 2.}},  {\em Phys.Rev.} {\bf 177} (1969)
  2247--2250.

\bibitem{Coleman:1969sm}
S.~R. Coleman, J.~Wess, and B.~Zumino, {\it {Structure of phenomenological
  Lagrangians. 1.}},  {\em Phys.Rev.} {\bf 177} (1969) 2239--2247.

\bibitem{Contino:2003ve}
R.~Contino, Y.~Nomura, and A.~Pomarol, {\it {Higgs as a holographic
  pseudoGoldstone boson}},  {\em Nucl.Phys.} {\bf B671} (2003) 148--174,
  [\href{http://xxx.lanl.gov/abs/hep-ph/0306259}{{\tt hep-ph/0306259}}].

\bibitem{Agashe:2004rs}
K.~Agashe, R.~Contino, and A.~Pomarol, {\it {The Minimal composite Higgs
  model}},  {\em Nucl.Phys.} {\bf B719} (2005) 165--187,
  [\href{http://xxx.lanl.gov/abs/hep-ph/0412089}{{\tt hep-ph/0412089}}].

\bibitem{Thaler:2005kr}
J.~Thaler, {\it {Little technicolor}},  {\em JHEP} {\bf 0507} (2005) 024,
  [\href{http://xxx.lanl.gov/abs/hep-ph/0502175}{{\tt hep-ph/0502175}}].

\bibitem{Grinstein:2011qm}
B.~Grinstein, R.~Kelley, and P.~Uttayarat, {\it {Hidden Fine Tuning In The
  Quark Sector Of Little Higgs Models}},  {\em PoS} {\bf ICHEP2010} (2010) 392,
  [\href{http://xxx.lanl.gov/abs/1102.4010}{{\tt arXiv:1102.4010}}].

\bibitem{Vecchi:2013bja}
L.~Vecchi, {\it {The Natural Composite Higgs}},
  \href{http://xxx.lanl.gov/abs/1304.4579}{{\tt arXiv:1304.4579}}.

\bibitem{Schmaltz:2010ac}
M.~Schmaltz, D.~Stolarski, and J.~Thaler, {\it {The Bestest Little Higgs}},
  {\em JHEP} {\bf 1009} (2010) 018,
  [\href{http://xxx.lanl.gov/abs/1006.1356}{{\tt arXiv:1006.1356}}].

\bibitem{Rao:2012gf}
K.~Rao and D.~Whiteson, {\it {Triangulating an exotic T quark}},  {\em
  Phys.Rev.} {\bf D86} (2012) 015008,
  [\href{http://xxx.lanl.gov/abs/1204.4504}{{\tt arXiv:1204.4504}}].

\bibitem{Schmaltz:2008vd}
M.~Schmaltz and J.~Thaler, {\it {Collective Quartics and Dangerous Singlets in
  Little Higgs}},  {\em JHEP} {\bf 0903} (2009) 137,
  [\href{http://xxx.lanl.gov/abs/0812.2477}{{\tt arXiv:0812.2477}}].

\bibitem{Cheng:2003ju}
H.-C. Cheng and I.~Low, {\it {TeV symmetry and the little hierarchy problem}},
  {\em JHEP} {\bf 0309} (2003) 051,
  [\href{http://xxx.lanl.gov/abs/hep-ph/0308199}{{\tt hep-ph/0308199}}].

\bibitem{Cheng:2004yc}
H.-C. Cheng and I.~Low, {\it {Little hierarchy, little Higgses, and a little
  symmetry}},  {\em JHEP} {\bf 0408} (2004) 061,
  [\href{http://xxx.lanl.gov/abs/hep-ph/0405243}{{\tt hep-ph/0405243}}].

\bibitem{Low:2004xc}
I.~Low, {\it {T parity and the littlest Higgs}},  {\em JHEP} {\bf 0410} (2004)
  067, [\href{http://xxx.lanl.gov/abs/hep-ph/0409025}{{\tt hep-ph/0409025}}].

\bibitem{Chang:2003un}
S.~Chang and J.~G. Wacker, {\it {Little Higgs and custodial SU(2)}},  {\em
  Phys.Rev.} {\bf D69} (2004) 035002,
  [\href{http://xxx.lanl.gov/abs/hep-ph/0303001}{{\tt hep-ph/0303001}}].

\bibitem{Chang:2003zn}
S.~Chang, {\it {A 'Littlest Higgs' model with custodial SU(2) symmetry}},  {\em
  JHEP} {\bf 0312} (2003) 057,
  [\href{http://xxx.lanl.gov/abs/hep-ph/0306034}{{\tt hep-ph/0306034}}].

\bibitem{Agashe:2006at}
K.~Agashe, R.~Contino, L.~Da~Rold, and A.~Pomarol, {\it {A Custodial symmetry
  for Zb anti-b}},  {\em Phys.Lett.} {\bf B641} (2006) 62--66,
  [\href{http://xxx.lanl.gov/abs/hep-ph/0605341}{{\tt hep-ph/0605341}}].

\bibitem{Peccei:1977hh}
R.~Peccei and H.~R. Quinn, {\it {CP Conservation in the Presence of
  Instantons}},  {\em Phys.Rev.Lett.} {\bf 38} (1977) 1440--1443.

\bibitem{Peccei:1977ur}
R.~Peccei and H.~R. Quinn, {\it {Constraints Imposed by CP Conservation in the
  Presence of Instantons}},  {\em Phys.Rev.} {\bf D16} (1977) 1791--1797.

\bibitem{Low:2002ws}
I.~Low, W.~Skiba, and D.~Tucker-Smith, {\it {Little Higgses from an
  antisymmetric condensate}},  {\em Phys.Rev.} {\bf D66} (2002) 072001,
  [\href{http://xxx.lanl.gov/abs/hep-ph/0207243}{{\tt hep-ph/0207243}}].

\bibitem{Gregoire:2003kr}
T.~Gregoire, D.~Tucker-Smith, and J.~G. Wacker, {\it {What precision
  electroweak physics says about the $SU(6)/Sp(6)$ little Higgs}},  {\em
  Phys.Rev.} {\bf D69} (2004) 115008,
  [\href{http://xxx.lanl.gov/abs/hep-ph/0305275}{{\tt hep-ph/0305275}}].

\bibitem{Hewett:2002px}
J.~L. Hewett, F.~J. Petriello, and T.~G. Rizzo, {\it {Constraining the littlest
  Higgs}},  {\em JHEP} {\bf 0310} (2003) 062,
  [\href{http://xxx.lanl.gov/abs/hep-ph/0211218}{{\tt hep-ph/0211218}}].

\bibitem{Csaki:2002qg}
C.~Csaki, J.~Hubisz, G.~D. Kribs, P.~Meade, and J.~Terning, {\it {Big
  corrections from a little Higgs}},  {\em Phys.Rev.} {\bf D67} (2003) 115002,
  [\href{http://xxx.lanl.gov/abs/hep-ph/0211124}{{\tt hep-ph/0211124}}].

\bibitem{Csaki:2003si}
C.~Csaki, J.~Hubisz, G.~D. Kribs, P.~Meade, and J.~Terning, {\it {Variations of
  little Higgs models and their electroweak constraints}},  {\em Phys.Rev.}
  {\bf D68} (2003) 035009, [\href{http://xxx.lanl.gov/abs/hep-ph/0303236}{{\tt
  hep-ph/0303236}}].

\bibitem{Kilian:2003xt}
W.~Kilian and J.~Reuter, {\it {The Low-energy structure of little Higgs
  models}},  {\em Phys.Rev.} {\bf D70} (2004) 015004,
  [\href{http://xxx.lanl.gov/abs/hep-ph/0311095}{{\tt hep-ph/0311095}}].

\bibitem{ATLAS:2012hpa}
{\bf ATLAS Collaboration} Collaboration, {\it {Search for exotic same-sign
  dilepton signatures ($b'$ quark, $T_{5/3}$ and four top quarks production) in
  4.7/fb of pp collisions at $\sqrt{s}=7$ TeV with the ATLAS detector}}, .

\bibitem{CMS:2012cya}
{\bf CMS Collaboration} Collaboration, {\it {Search for a heavy partner of the
  top quark with charge 5/3}}, .

\bibitem{Godfrey:2012tf}
S.~Godfrey, T.~Gregoire, P.~Kalyniak, T.~A. Martin, and K.~Moats, {\it
  {Exploring the heavy quark sector of the Bestest Little Higgs model at the
  LHC}},  {\em JHEP} {\bf 1204} (2012) 032,
  [\href{http://xxx.lanl.gov/abs/1201.1951}{{\tt arXiv:1201.1951}}].

\bibitem{Kearney:2013oia}
J.~Kearney, J.~Thaler, and A.~Pierce, {\it {Top Partner Probes of Extended
  Higgs Sectors}},  \href{http://xxx.lanl.gov/abs/1304.4233}{{\tt
  arXiv:1304.4233}}.

\bibitem{Cheung:2006ij}
C.~Cheung and J.~Thaler, {\it {(Reverse) engineering vacuum alignment}},  {\em
  JHEP} {\bf 0608} (2006) 016,
  [\href{http://xxx.lanl.gov/abs/hep-ph/0604259}{{\tt hep-ph/0604259}}].

\bibitem{ArkaniHamed:2001nc}
N.~Arkani-Hamed, A.~G. Cohen, and H.~Georgi, {\it {Electroweak symmetry
  breaking from dimensional deconstruction}},  {\em Phys.Lett.} {\bf B513}
  (2001) 232--240, [\href{http://xxx.lanl.gov/abs/hep-ph/0105239}{{\tt
  hep-ph/0105239}}].

\end{thebibliography}\endgroup
\bibliographystyle{JHEP}

\end{document}